\documentclass[12pt,journal,compsoc]{IEEEtran}
\usepackage{epsfig}
\usepackage{epstopdf}
\usepackage[T1]{fontenc}
\usepackage{amssymb}
\usepackage{amsmath}
\usepackage{amsfonts}
\usepackage{verbatim}
\usepackage{graphicx}
\usepackage{hyperref}
\usepackage{algorithmic}
\usepackage{algorithm}
\usepackage{array}
\usepackage{multirow}

\usepackage{pgfplots}

\usepackage{tikz-timing}[2009/05/15]
\ifCLASSOPTIONcompsoc
\else
\fi

\ifCLASSINFOpdf
\else
\fi

\begin{document}
\title{ Hardware Implementation of four\\ byte per clock RC4 algorithm}

\author{Rourab~Paul,
        Amlan~Chakrabarti,~\IEEEmembership{Senior Member,~IEEE,}
        and~Ranjan~Ghosh,% <-this % stops a space
\IEEEcompsocitemizethanks{\IEEEcompsocthanksitem M. Shell is with the Department
of Electrical and Computer Engineering, Georgia Institute of Technology, Atlanta,
GA, 30332.\protect\\
E-mail: see http://www.michaelshell.org/contact.html
\IEEEcompsocthanksitem J. Doe and J. Doe are with Anonymous University.}% <-this % stops a space
\thanks{Manuscript received April 19, 2005; revised January 11, 2007.}}

\markboth{Journal of \LaTeX\ Class Files,~Vol.~6, No.~1, January~2007}%
{Shell \MakeLowercase{\textit{et al.}}: Bare Demo of IEEEtran.cls for Computer Society Journals}

\IEEEcompsoctitleabstractindextext{%
\begin{abstract}
%\boldmath
In the field of cryptography till date the 2-byte in 1-clock is the best known RC4 hardware design \cite{ieee:two_byte}, while 1-byte in 1-clock \cite{springerlink:one_byte}, and the 1-byte in 3 clocks \cite{IEEE:b}\cite{patent:matthews} are the best known implementation. The design algorithm in\cite{springerlink:one_byte} considers two consecutive bytes together and processes them in 2 clocks. The design \cite{ieee:two_byte} is a pipelining architecture of \cite{springerlink:one_byte}. The design of 1-byte in 3-clocks is too much modular and clock hungry. In this paper considering the RC4 algorithm, as it is, a simpler RC4 hardware design providing higher throughput is proposed in which 6 different architecture has been proposed. In design 1, 1-byte is processed in 1-clock, design 2 is a dynamic KSA-PRGA architecture of Design 1. Design 3 can process 2 byte in a single clock, where as Design 4 is Dynamic KSA-PRGA architecture of Design 3. Design 5 and Design 6 are parallelization architecture design 2 and design 4 which can compute 4 byte in a single clock. The maturity in terms of throughput, power consumption and resource usage, has been achieved  from design 1 to design 6. The RC4 encryption and decryption designs are respectively embedded on two FPGA boards as co-processor hardware, the communication between the two boards performed using Ethernet. 
\end{abstract}
% not want either math or citations to appear in the abstract.

}

\maketitle

% is not selected <OR> if conference mode is selected - because compsoc
% conference papers position the abstract like regular (non-compsoc)
\IEEEdisplaynotcompsoctitleabstractindextext

% \fi
\IEEEpeerreviewmaketitle

\section{Introduction}

\IEEEPARstart{R}{C4} is a widely used stream cipher whose algorithm is very simple. It has withstood the test of time in spite of its simplicity. The RC4 was proposed by Ron Rivest in 1987 for RSA Data Security and was kept as trade secret till 1994 when it was leaked out \cite{DBLP:spaul}. Today RC4 is a part of many network protocols, e.g. SSL, TLS, WEP, WPA and many others. There were many cryptanalysis to look into its key weaknesses\cite{DBLP:spaul} \cite{springerlink:gpaul} followed by many new stream ciphers \cite{t:good} \cite{p:leg} . RC4 is still the popular stream cipher since it is executed fast and provides high security.\\(Reference needed) 
	There exist hardware implementations of some of the stream ciphers in the literature \cite{p:kitsos} \cite{m:gal} \cite{dp:math}. Since about 2003 when FPGA technology has been matured to provide cost effective solutions, many researchers started hardware implementation of RC4 as a natural fall out \cite{IEEE:b} \cite{patent:matthews}.  The FPGA technology turns out to be attractive since it provides soft core processor having design specific functional capability of a main processor \cite{xilinx:online} along with reconfigurable logic blocks that can be synthesized to a desired custom coprocessor, embedded memories and IP cores. One can design RC4 algorithm totally as an executable code for the soft core processor (main processor) only or in custom coprocessor hardware operated by the main processor. Because of the system overhead, any single instruction if executed in the main processor takes at least 3 clocks, while the identical one when executed in a coprocessor takes 2 or 1-clock as the latter is customized to handle the specific task. Besides the clock advantage, the coprocessor based design makes the system throughput faster by another fold since it is executed in parallel with the main processor.\\
	In this paper, RC4 algorithm is considered as it is and exploiting conventional VHDL features a design methodology is proposed processing of RC4 algorithm in two different hardware approaches. 1st architecture named as design 1 and design 2, which can execute 1 byte in single clock, the 2nd approach can execute 2 bytes in single clock which is named as design 3 and design 4 and lastly these two architectural  approaches (design 2 and design 4) has been accommodated in coprocessor environment named as design 5 and design 6 to increase the system throughput. The said design is implemented in a custom coprocessor functioning in parallel with a main processor (Xilinx Spartan3E XC3S500e-FG320 and Virtex5 LX110t  FPGA architecture) followed by secured data communication between two FPGA boards through their respective Ethernet ports $–$  each of the two boards performs RC4 encryption and decryption engines separately. The performance of our design in terms of number of clocks proved to be better than the previous works \cite{ieee:two_byte}, \cite{springerlink:one_byte}, \cite{IEEE:b}, \cite{patent:matthews}. The dynamic KSA-PRGA architecture(design 3 and design 4) is introduced to reduce system power and resource utilization. 
\subsection{Existing Work}
In the year 2004 P. Kitsos et al \cite{IEEE:b} has tried a 3 byte per clock architecture of RC4 algorithm. At the first clock of PRGA the $i$ and $j$ is computed. In next clock $S[i]$ and $S[j]$ has been extracted from RAM and added. At the same step the addition  has been stored in register $t$. At third clock swapping process has been done as well the value of $t$ register used to address the key $S[t]$.  The swapping and Z calculation is going on in the 3rd clock. When the Z is being computed $t$ can not address the value at the swapped s-box, because swapped s-box will be updated after a clock pulse. This timing distribution of task can give error while $S[i]+S[j]=i~or~j$.\\
In 2003 Matthews, Jr. \cite{patent:matthews} has proposed another 3 byte clock architecture where KSA and PRGA both have 3 clock for each iteration. On 2008 Matthews again proposed a new 1 byte per clock pipelined architecture in article Jr.\cite{dp:math}. Lastly S. Sen Gupta et al has proposed 2 architectures, such that 1 byte per clock \cite{springerlink:one_byte} and 2 byte per clock \cite{ieee:two_byte} architecture. Article \cite{springerlink:one_byte} is an excellent loop unrolled architecture where 2 byte is executing in 2 consecutive clocks. Article \cite{ieee:two_byte} is pipelined architecture of 1 byte clock architecture where 4 byte can be executed in 2 clocks. 

\subsection{Our Contribution}
We have proposed 6 different architectures for RC4 algorithm to increase the throughput compromising with resource and power consumption and trying to search this most optimized architecture in terms of throughput, resource usage and power. The main contribution of this paper is described as a various design approaches.\\

\textbf{Design 1}: We have designed a RC4 architecture with 1 byte per clock throughput. This design has not been done using any loop unrolling method. It is a dual edge sensitive clock  architecture, incorporate with pipelining concept.\\ 

\textbf{Design 2}: This design is identical with the design 1 which was consuming more power and more resource to have 1 byte per clock throughput. Keeping these points in mind we were searching some optimizing technique. We noticed that the KSA and PRGA processes are very much identical to each other. In design 1 when the PRGA starts the KSA does not have any significance of occupying the FPGA dice according to the algorithm concern. The existence of both of the PRGA and KSA  process causes large silicon area leading with more power consumption  but in recent days the FPGA vendors like XILINX, ALTERA does not have any provision to cut down the $V_{cc}$ voltage line of those MOSFETs which are unnecessary after its execution like KSA process. If the said provision can be accommodated on FPGA platform it can save large power consumption and resources. Recently XILINX has launched a partial reconfiguration tool which can modify hardware in runtime but without this tool we can use some smart trick to modify the KSA process into PRGA dynamically. This means the same hardware can be utilized by PRGA process, previously which was used by KSA process. In this design 2 we have introduced some logic with the algorithm which can transform the KSA process(after its execution) into PRGA process. The design 2 architecture is named as Dynamic KSA-PRGA architecture(DKP architecture).  Design 2 has reduced huge power consumption as well as resource usage without compromising of throughput. The results of power consumption and resource usage supporting our claim.\\

\textbf{Design 3}: To increase The throughput of this architecture we used a loop unrolling method which is mainly motivated from \cite{springerlink:one_byte} but having a unique different hardware approach. This design 3 architecture can execute 2 byte in a single clock which means, comparing with the previous designs, the throughput is doubled without contributing large power and usage.\\  

\textbf{Design 4}: Design 4 is a DKP form of design 3 to save the wattage and slice-LUTs.\\

\textbf{Design 5}: To have more throughput we used 4 parallel co-processor where each is coprocessor executing the design 2 architecture, parallel with a main processor. The main processor is responsible for data receiving and transmitting part. The data has been accused from PS2 port, RS232 port and Ethernet port. Only the 4 co-processors can be accommodate for parallel processing, because the data bus of microblaze processor is 32 bit, so coprocessors can communicate only 4 keys(1 key from each co-processor) with main processor at a time.\\

\textbf{Design 6}: In design 6 we used 2 parallel co-processor where each coprocessor is executing the design 4 architecture, parallel with a main processor. Only the 2 co-processor can be accommodate for parallel processing, because the data bus of microblaze processor is 32 bit, so co-processors can communicate only 4 keys(2 key from each co-processor) with main processor at a time. The other hardwares of design 6 is similar to design 5. \\
%I wish you the best of success.
 
\hfill January 11, 2007

%\subsection{Subsection Heading Here}
%Subsection text here.
%\begin{figure}[htbp]
%\centering
%\includegraphics[width=9cm,height=4cm]{./figure/dynamic.eps}
%\rule{15em}{0.1pt}
%\caption[An Electron]{An electron (artist's impression).}
%\label{fig:Electron}
%\end{figure}

%\IEEEpubidadjcol

\section{RC4 Algorithm}\label{rc4algo}
RC4 has a S-Box S[N], N = 0 to 255 and a secret key, key[l] where l is typically between 5 and 16, used to scramble the S-Box [N]. It has two sequential processes, namely KSA (Key Scheduling Algorithm) and PRGA (Pseudo Random Generation Algorithm) which are stated below in figure \ref{fig:ksa_prga_fig}.

\begin{figure}[!htb]
\centering
\vspace{-9pt}
\includegraphics[width=9cm, height=4.5cm]{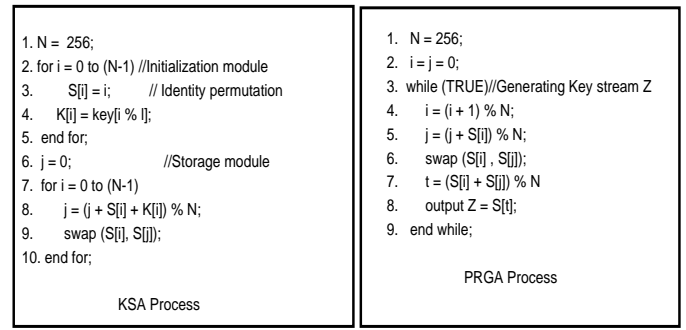}
\vspace{-6pt}
\caption{RC4 Algorithm}
\vspace{-18pt}
\label{fig:ksa_prga_fig}
\end{figure}

\section{Hardware Implementation of 1-byte 1-clock design (Design 1 and Design 2)}
\label{design12}
\begin{figure}[htbp]
\centering
\vspace{-9pt}
\includegraphics[width=9cm, height=8cm]{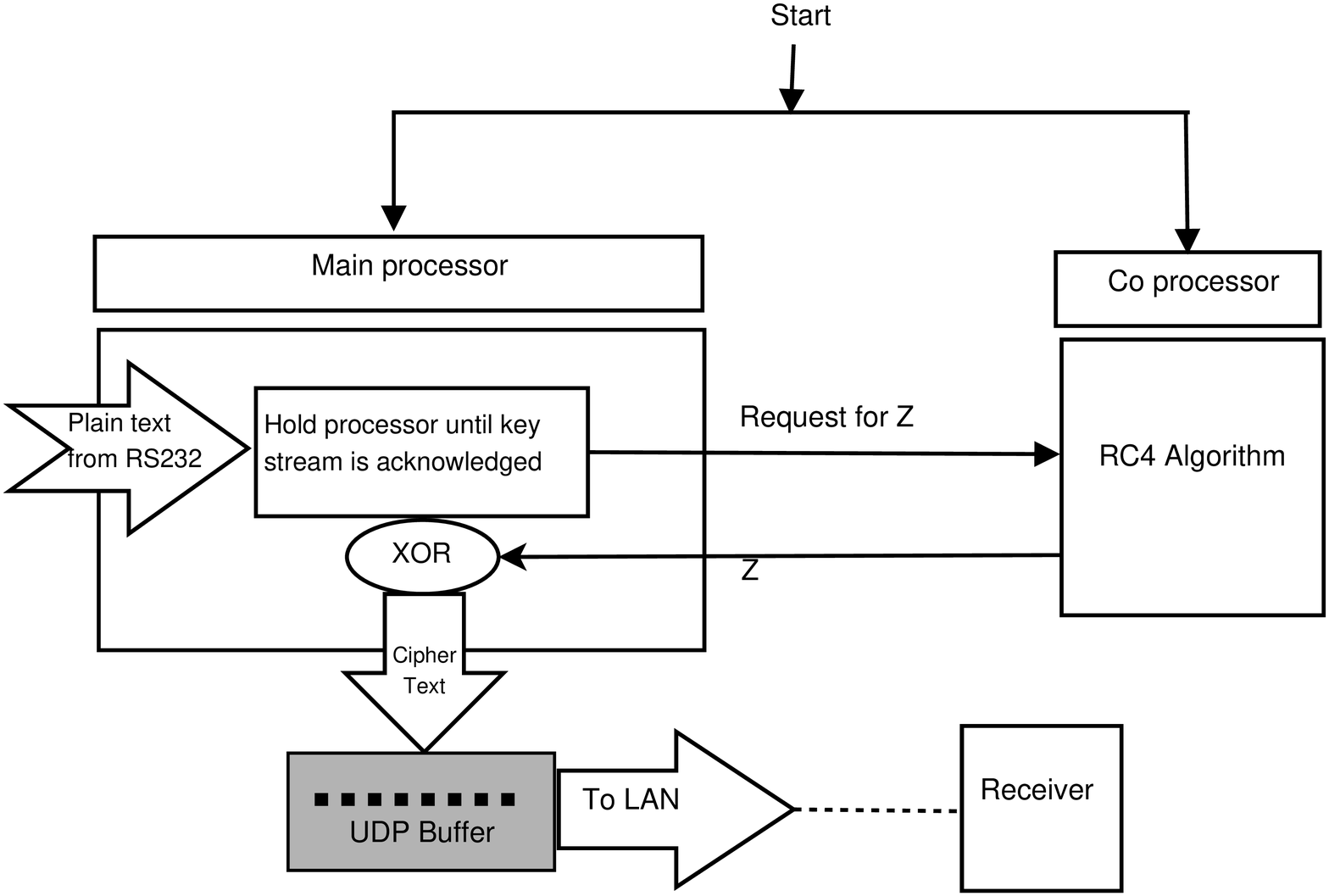}
\vspace{-6pt}
\caption{Functioning of the main processor with the co processor}
\vspace{-8pt}
\label{fig:co_pro}
\end{figure}

Figure \ref{fig:co_pro} shows key operations performed by the main processor in conjunction with a coprocessor till the ciphering of the last text character. The hardware for realizing RC4 algorithm comprises of KSA and PRGA units, which are designed in the coprocessor as two independent units, and the XOR operation is designed to be done in the main processor. The central idea of the present embedded system implementation of one RC4 byte in 1-clock is the hardware design of a storage block shown in Figure \ref{fig:s_box}, which is used in the KSA as well as in the PRGA units. The storage block contains a common S-Box connected to dual select MUX-DEMUX combination and executes the swap operation following line 9 of Algorithm 1 and line 6 of Algorithm 2, in order to update the S-Box. The swap operation in hardware is explained in the following sub-section.\\

\subsection{Storage Block Updating the S-Box}
\begin{figure}[htbp]
\centering
\vspace{-1pt}
\includegraphics[width=9cm]{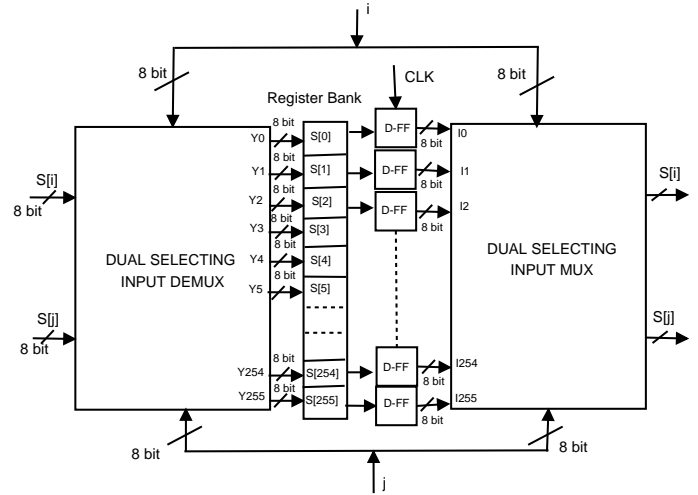}
\vspace{-6pt}
\caption{Storage Block updating the S-Box}
\vspace{-1pt}
\label{fig:s_box}
\end{figure}

The storage block consists of a register bank containing 256 numbers of 8-bit data representing the S-Box(register bank), 256:2 MUX, 2:256 DEMUX and 256 D flip-flops. Each of the MUX and DEMUX is so designed, that with 2-select inputs $i, j$ (8 bit width) can address the two register data $S[i] and S[j]$  at the same time (The VHDL complier has merged 2 256:1 MUX to design this single 256:2 MUX, The same thing is happened for 2:256 DEMUX). The hardware design of the storage block updating the S-Box is shown in figure \ref{fig:s_box} . For swapping, the same S-Box is accessed by the KSA unit with its MUX0-DEMUX0 combination and also by the PRGA unit with its MUX2-DEMUX2 combination and it is also accessed by another MUX3 in PRGA for the generation of the key stream Z. The storage block swaps S[i] and S[j] and thereby updates the S-Box. For swapping, S[i] and S[j] ports of MUX are connected to S[j] and S[i] ports of DEMUX respectively. This storage block has thus 3 input ports (i, j and CLK), and 2 inout ports (S[i] of MUX, S[j] of DEMUX and S[j] of MUX, S[i] of DEMUX). The storage block of PRGA unit provides 2 output ports from its 2 inout ports which are fed to an adder circuit with MUX3. During the falling edge of a clock pulse, S[i] and S[j] values corresponding to ith and jth locations of the register bank are read and put on hold to the respective D flip-flops. During the rising edge of the next clock pulse, the S[i] and S[j] values are transferred to the MUX outputs and instantly passed to the S[j] and S[i] ports of the DEMUX respectively and in turn are written to the jth and ith locations of the register bank. The updated S-Box is ready during the next falling edge of the same clock pulse, if called for.\\

\subsection{Design of the KSA Unit following Algorithm}
\begin{figure}[!htb]
\centering
\vspace{-6pt}
\includegraphics[width=9cm]{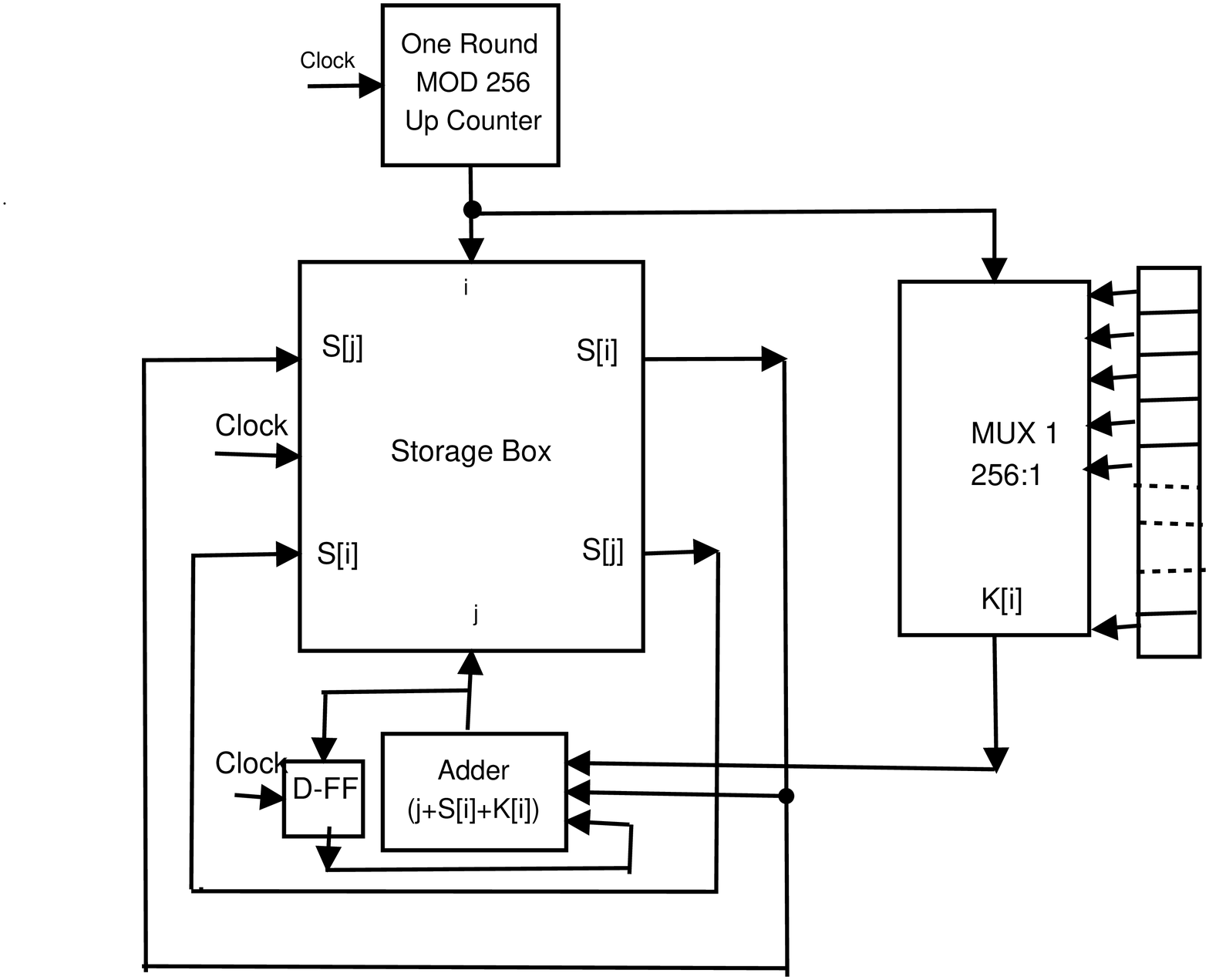}
\vspace{-6pt}
\caption{Schematic Design of the KSA Unit}
\vspace{-6pt}
\label{fig:ksa}
\end{figure}

Figure \ref{fig:ksa} shows a schematic diagram of design of the KSA unit. Initially the S-Box is filled with identity permutation of $i$ whose values change from 0 to 255 as stated in line 3 of initialization module.  The l-bytes of secret key are stored in the K[256] array as given in line 4. The KSA unit does access its storage block with i being provided by a one round of MOD 256 up counter, providing fixed 256 clock pulses and j being provided by a 3-input adder (j, S[i], and K[i]) following the line 8 of the storage module, where j is clock driven, S[i] is MUX0 driven chosen from the S-Box and K[i] is MUX1 driven chosen from the K-array. The S-Box is scrambled by the swapping operation stated in line 9 using MUX0-DEMUX0 combinations in the storage block. The KSA operation takes one initial clock and subsequent 256 clock cycles. 

\subsection{Design of the PRGA Unit following Algorithm}
\begin{figure}[!htb]
\centering
\vspace{-6pt}
\includegraphics[width=9cm, height=9cm]{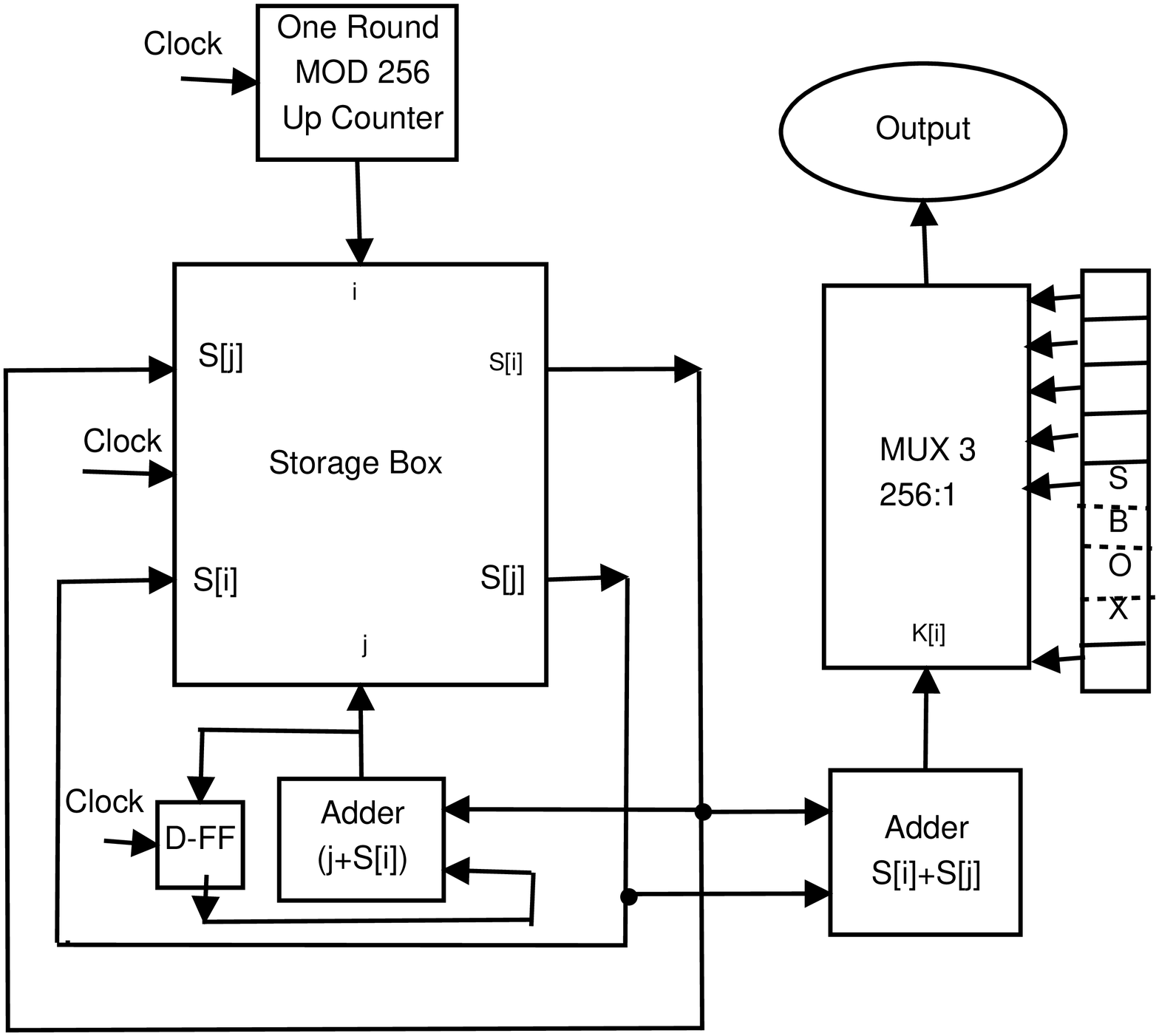}
\vspace{-6pt}
\caption{Schematic Design of the PRGA Unit}
\vspace{-6pt}
\label{fig:prga}
\end{figure}

Figure \ref{fig:prga} shows a schematic diagram of  the  design of the PRGA  unit. The PRGA unit does access the storage block with i being provided by a MOD 256 up counter (line 4) and j being given by a 2-input (j and S[i]) adder  following the line 5 where j is clock driven and S[i] is MUX2 driven chosen from the S-Box. With updated j and current i, the swapping of S[i] and S[j] is executed following line 6 using MUX2-DEMUX2 combination of the storage block. Following the line 7, S[i] and S[j] give a value of t based on which the key stream Z is selected from the S-Box using MUX 3 (line 8).

\subsection{Timing analysis of the PRGA Operation}
\begin{figure}[!htb]
\centering
\vspace{-6pt}
\includegraphics[width=9cm, height=5cm]{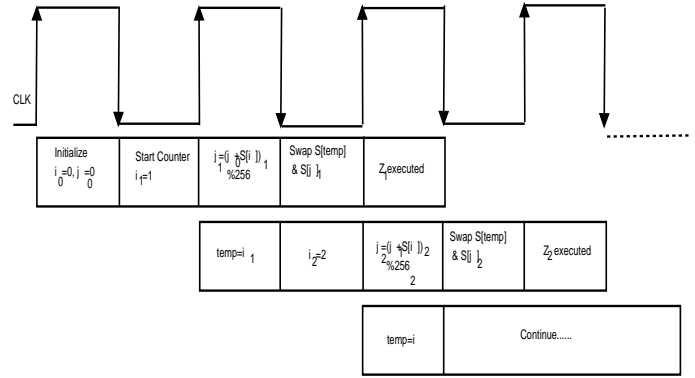}
\vspace{-6pt}
\caption{Clock-wise description of PRGA operation}
\vspace{-6pt}
\label{fig:prga_timing}
\end{figure}

Let $\phi_i$ denotes the $i^{th}$ clock cycle for $i \geq 0$. It is assumed that the PRGA unit starts when clock cycle is $\phi_0$. It is observed that a signal value gets updated during a falling edge of a clock cycle if it is changed during the rising edge of the previous clock cycle. The symbol $ \leftrightarrow$ indicates swap operation. The clock-wise PRGA operations are shown in Figure \ref{fig:prga_timing}.

\textbf{Timing Analysis of PRGA}\\
The MOD 256 up counter shown in Figure \ref{fig:prga} is so designed that i starts from 1, goes up to 255 and then it repeats from 0 to 255 for each 256 subsequent clock cycles.\\
\begin{itemize}

\item Rising edge of  $\phi_0$: Initialize $j_0$=0 and $i_0$=0.\\
\item Falling edge of $\phi_0$: Start counter $i_1$ =1.\\
\item Rising edge of  $\phi_1$: $j_1$=(j0 + S[i1]) \%\ 256; temp=$i_1$.\\
\item Falling edge of $\phi_1$: $i_2 \rightarrow$2;  S[temp]  $\leftrightarrow S[j_1]$;\\
\item Rising edge of  $\phi_2: j_2=(j_1+S[i_2]) \%\ 256;$ temp=$i_2$, $Z_1=(S[i_1]+S[j_1])$ \%\ 256 = 1st Key.\\
\item Falling edge of $\phi_2$: $i_3 \rightarrow$3;  S[temp] $\leftrightarrow$S[$j_2$]; \\
\item Rising edge of $\phi_3$: $j_3=(j_2+S[i_3]) \%256,$ temp=$i_3$, $Z_2=(S[i_2]+S[j_2])$ \%\ 256 = 2nd Key stream. \\
\end{itemize}
%j2=(j1+S[i2]) \%\ 256; Read S[i2], S[j2]. \\
%Rising edge of  $\phi_2$: S[i2]$ \leftrightarrow$S[j2]; Z2=(S[i2]+S[j2]) \%\ 256 = 2nd Key stream. \\
%Falling edge of $\phi_2$: i3$ \rightarrow$3; j3=(j2+S[i3]) \%\ 256; Read S[i3], S[j3].\\
The series continues generating successive keys (Z$’$s). If the text characters are n and n > 254, i = 0 after the first round and the clock repeats (n+2) times. After generating $Z_n$ during the rising edge of $\phi_{n+1}$, PRGA stops. For generating Zn, the PRGA requires (n+2) clocks and its throughput per byte is (1+2/n). \\
\subsection{Dynamic KSA-PRGA Architecture (Design 2)}
If we look into the KSA and PRGA process in Figure \ref{fig:ksa_prga_fig} we can find both of these processes are almost identical. Line 8,9 of KSA process and line 5, 6 of PRGA process are selfsame, except the extra addend secret key (K[i]) in KSA process. The key execution(Z) is also an additional process in PRGA. In section 3  design we have used two separate sequential blocks for both KSA and PRGA. As these processes are separated and executed sequentially, the algorithm core is contributing few extra static power, and dynamic power with some additional resource usage.  To overcome these flaws we are proposing a single process dynamic KSA-PRGA architecture where KSA process itself is transformed to PRGA process after 257 clocks. Due to the reutilization of LUTs, slices and FF-pairs the proposed architecture can reduce significant resource usage which has led to low power design. \\
\begin{figure}[htb]
\centering
\vspace{-2pt}
\includegraphics[width=9cm, height=9cm]{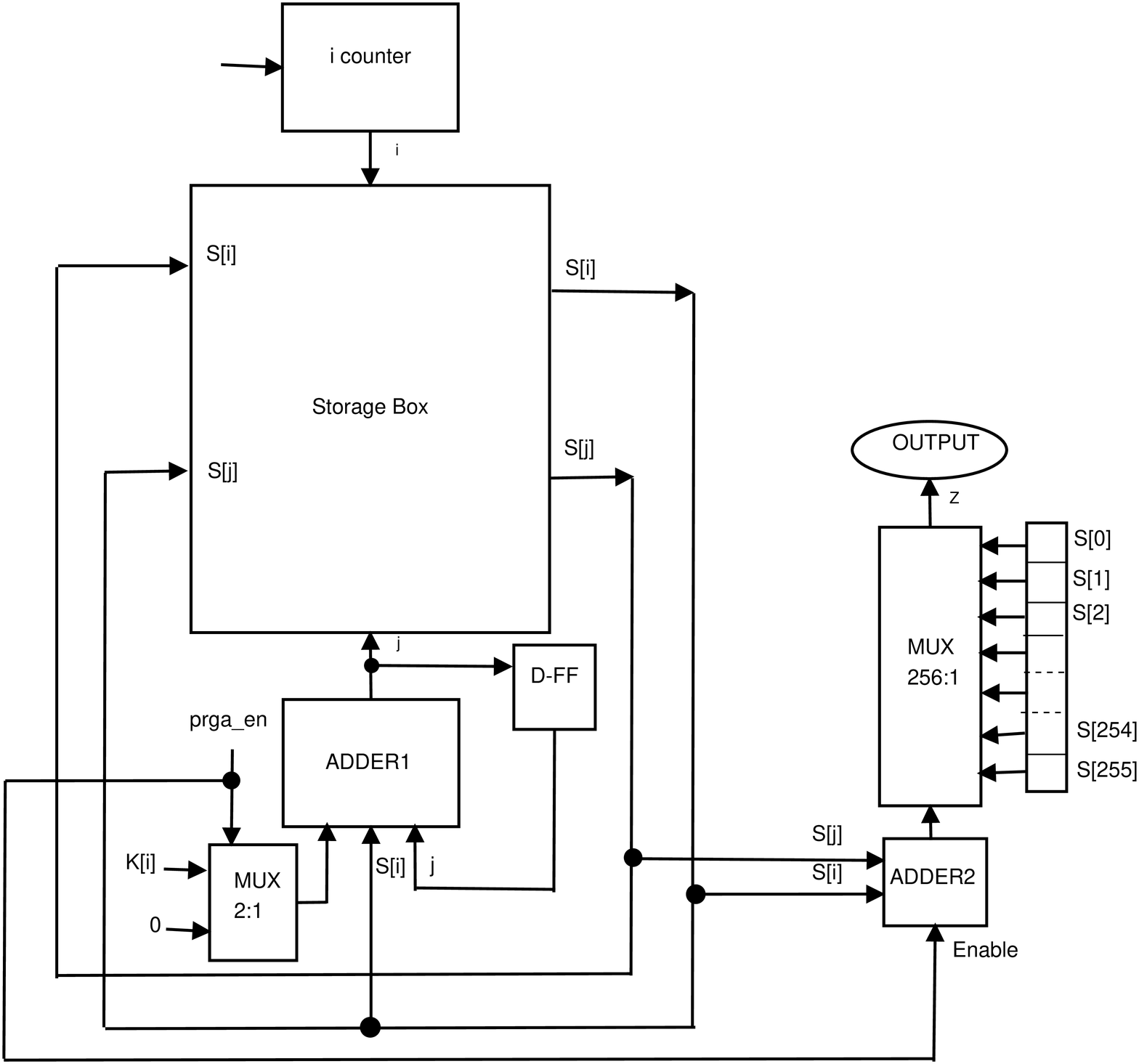}
\vspace{-2pt}
\caption{Dynamic KSA-PRGA}

\label{fig:dynamic}
\end{figure}

\subsubsection{~~~~~Hardware Overview}
For the dynamic KSA-PRGA architecture few additional hardware has been added and some significant alteration of previous normal structural design has been done. The whole changes has been described below,\\
Instead of using two separated $ 'i-counter' $ for KSA and PRGA process we have used a single $ 'i-counter' $ in our new architecture. Previous $ 'i-counter' $ for KSA is a single round MOD-256 up counter and PRGA $ 'i-counter' $ was continuous MOD-256 up counter which has a counting range of  0 to 255 skipping the initial 0 count in first round. For both of these processes the dynamic architecture uses a simple continuous MOD-256 up counter which starts counting  from count 0 and ends with 255 to finish the entire KSA process. After finishing the KSA process the counter is again reset to 0 count. At this 0 count the PRGA process is initialized (Initialized j by 0) and from the next successive count it starts the key execution process.
A signal named as $prga~enable (prga\_en)$ is employed to detect the KSA process status. When KSA has been completed $prga~enable$ has been latched high. During the KSA process $prga~enable$ stands 0 logic state to pass the $K[i]$ to Adder1 through the 2:1 MUX but when $prga~enable$ goes high, $0$ value will be forwarded to Adder1. So, at that time the Adder1 is started to execute $(j + S[i])$ instead of $(j + S[i] + K[i])$ as stated in PRGA algorithm. At the same time the $prga~enable$ signal enables the Adder2 circuit to execute the Z(keys). These two activation on main hardware block has altered the KSA process to PRGA process. Figure \ref{fig:dynamic} shows the architectural block diagram of dynamic KSA-PRGA. Resource usage and power consumption of dynamic KSA-PRGA architecture has been compared with previous architecture in table \ref{table:dynamic_usage} and \ref{table:dynamic_power} and Figure \ref{fig:power_graph} is the graphical representation of table \ref{table:dynamic_usage} and \ref{table:dynamic_power}.

\section{Hardware Implementation of 2-byte 1-clock design (Design 3 and Design 4)}
\label{design34}
This approach is mainly motivated from reference [saurav senguptas paper] which is based on loop unrolling method.
%---------------------------------------------------------------------------------
\begin{table}[]
\caption{Two unrolled loops} % title of Table
\centering  % used for centering table
\resizebox{9cm}{!}{%

    \begin{tabular}{|c|c|c|  }
        \hline
Steps& 1st iteration & 2nd iteration\\\hline
1& $i_1 = i_0 + 1$ & $i_2 = i_1 + 1 = i_0 + 2$\\\hline
2&$j_1 = j_0 + S_0[i_1]$&$j_2 = j_1 + S_1[i_2]$\\
~&~&$ = j_0 + S_0[i_1] + S_1[i_2]$\\\hline
3&$Swap S_0[i_1]\leftrightarrow S_0[j_1]$&$Swap S_1[i_2] \leftrightarrow S_1[j_2]$\\\hline
4&$Z_1 = S_1[S_0[i_1] + S_0[j_1]]$&$Z_2 = S_2[S_1[i_2] + S_1[j_2]]$\\
\hline
    \end{tabular}
}
\label{table:2steps_table} % is used to refer this table in the text
\end{table}
%--------------------- end of the example ------------------------
\subsection{Storage Block Updating the S-Box}
\begin{figure}[!htb]
\centering
\vspace{-6pt}
\includegraphics[width=9cm]{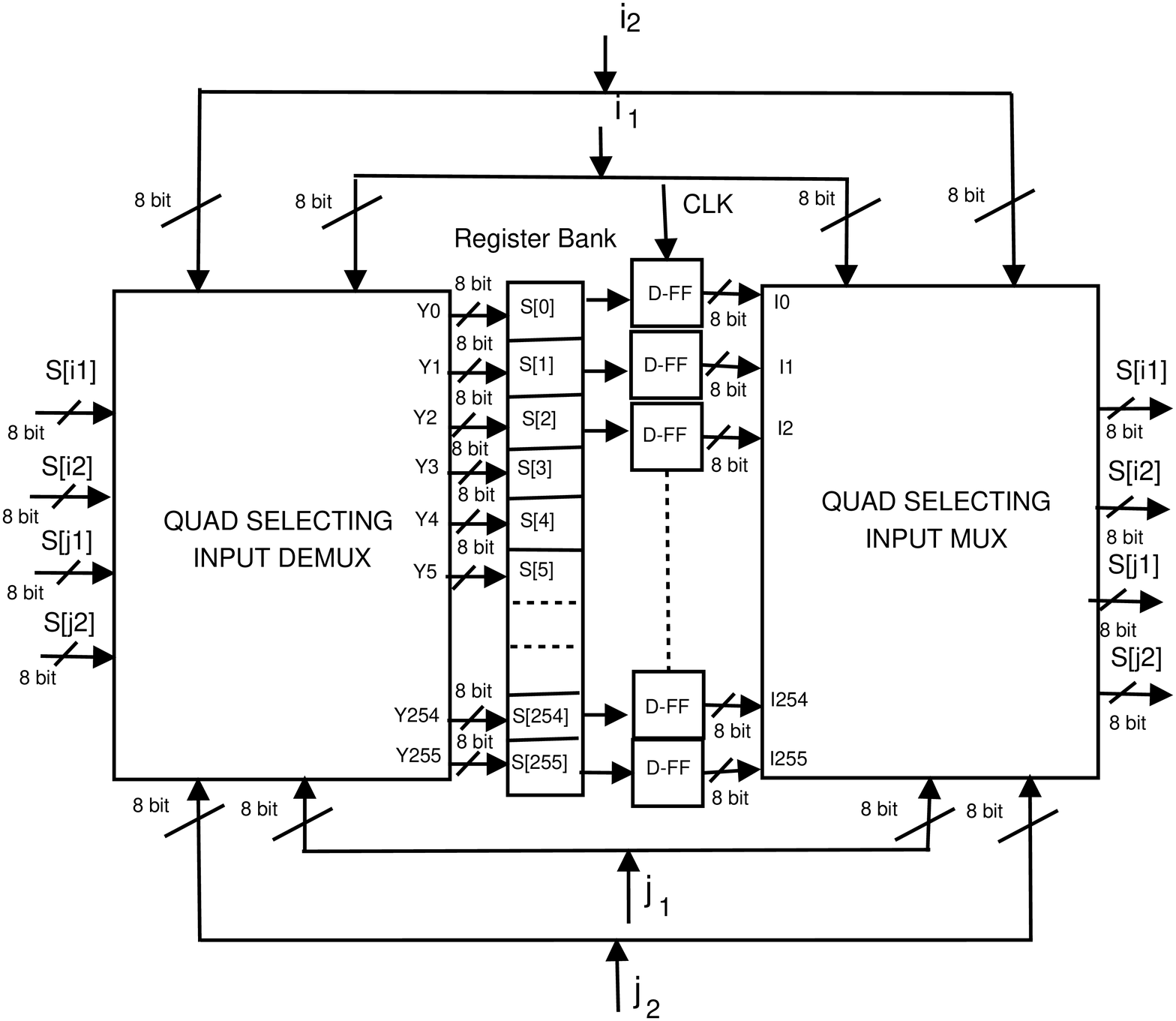}
\vspace{-6pt}
\caption{Storage Block updating the S-Box}
\vspace{-8pt}
\label{fig:s_box2}
\end{figure}

Figure \ref{fig:s_box2} shows a schematic diagram of design of the storage box unit which is identical with figure \ref{fig:s_box}. The only difference between the s-box of 1 byte 1 clock architecture and s-box of 2 byte 1 clock architecture is number of ports. This storage block also consists of a register bank containing 256 numbers of 8-bit data representing the S-Box (register bank), 256:1 MUX, 1:256 DEMUX and 256 D flip-flops.  Here two sets of $i$ and $j$ can address 4 s-box element at a time. The value of $S[i_1], S[i_2], S[j_1]$ and $S[j_2]$ are updating by a 4 input DEMUX followed by a 4 input MUX via Swap Controlling block.
\subsection{KSA unit of 2 byte per clock architecture}
\subsubsection{ ~~~~$j_1$ and $j_2$ Generation of KSA unit}
\begin{figure}[!htb]
\centering
\vspace{-6pt}
\includegraphics[width=9cm, height=6 cm]{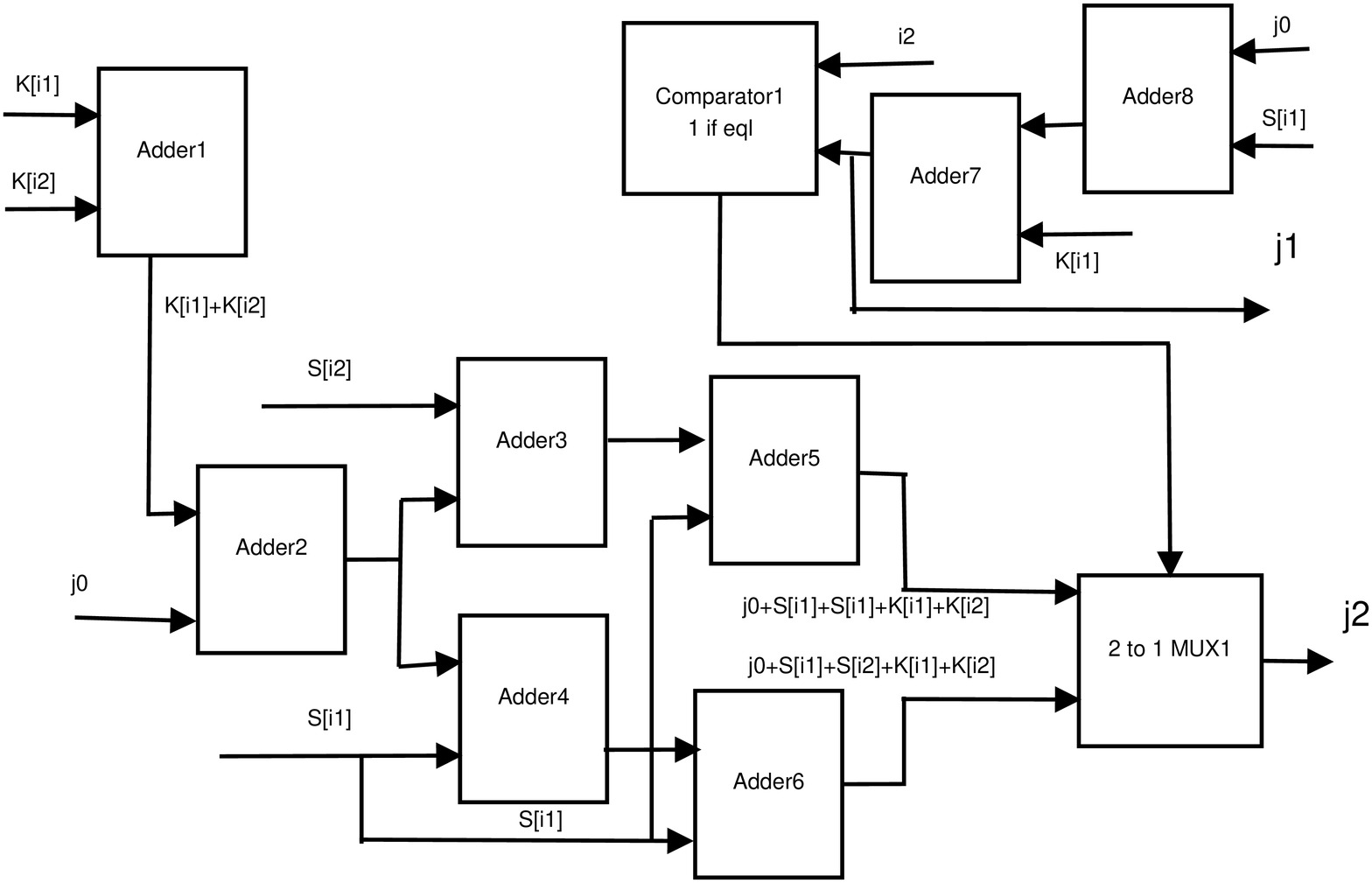}
\vspace{-6pt}
\caption{j1 and j2 genarator of KSA unit}
\vspace{-8pt}
\label{fig:j1_j2_ksa_fig}
\end{figure}

In this proposed architecture we need to compute $j_1$ and $j_2$ at the same clock instance. $j_1$ computation circuit is same as the previous KSA architecture shown in  \ref{fig:ksa} but in the $j_2$ circuitry we need to adopt few tricks to compute the $j_2$ at that same clock while $j_1$ is executing. At the j computation $j_0$ has been initiated by $0$ value. According to the conventional RC4 algorithm $j_1$ is executed at the step 2 of table \ref{table:2steps_table}. In figure \ref{fig:j1_j2_ksa_fig} Adder7 and Adder 8 is responsible to add $j_0+S_0[i_1]+K[i_1]$ to evaluate the value of $j_1$. For the  $j_2$ computation we again need to see 1st column step2 of table \ref{table:2steps_table} where the $j_2$ computation may be divided into the following two cases\\
\begin{equation}
\resizebox{.9\hsize}{!}{ $j_2=j_0+S_{0}[i_1]+S_{1}[i_2]+2 keys\left\{\begin{array}{ll}\mbox{$ j_0+S_{0}[i_1]+S_{0}[i_2]+2keys$} & \mbox{if $i_2 \neq j_1$};\\
\mbox{$j_0+S_{0}[i_1]+S_{0}[i_1]+2keys$} &  \mbox{$i_2 =j_1$}.\\
\end{array}\right. $}
\end{equation}
It has been noted that the only differentiation from $S_0$ to $S_1$ is the swap process, so it is very obvious to check if $i_2$ is equal to either of $i_1$ or $j_1$. As $i_2$ never be equal to $i_1$ (differ only by 1 modulo 256) we only need to check $i_2$ and $j_1$ which will decide whether $S_{0}[i_2]$  or $S_{0}[i_2]$ will be passed as a 3rd operand of addition process. Figure \ref{fig:j1_j2_ksa_fig} shows the block diagram of $j_1$ and $j_2$ generator. Adder 1 is adding the two successive secret keys which is addressed by $i_1$ and $i_2$. Adder 2 is summing $k[i_1]+K[i_2]$ with $j_0$ and passing this value to Adder3 and adder4. Adder3 and adder4 is adding $k[i_1]+K[i_2]+j_0$ with $S[i_2]$ and $S[i_1]$ respectively. Adder5 and Adder 6 is responsible to execute $j_0+S_0[i_1]+S_0[i_2]+K[i_1]+K[i_2]$ and $j_0+S_0[i_1]+S_0[i_1]+K[i_1]+K[i_2]$.

\subsubsection{~~~~~Swap Controlling block}In line 9 of figure \ref{fig:ksa_prga_fig} a swapping process is updating s-box in each clock cycle. In 2 byte per clock architecture 2 loop iteration has been unrolled in a single loop iteration. Here two successive swap process should be accumulated in a single iteration.  To replace this two swap processes by a single swap process we need to check all possible relation of $i_1, i_2, j_1$ and $j_2$. We need to check if $i_2$ and $j_2$ can be equal to $i_1$ or $j_1$. Among the all possible correlations we can skip the combination of $i_1$ and $i_2$ as $i_2=i_1+1$. For the remaining 3 cases, such that $i_2$ with $j_1$,  $j_2$ with $i_1$ and $j_2$ with $j_1$ we need 3 comparator  circuits (8 bit) to compare each of their relation in every clock cycle. Table \ref{table:swap_table} is showing the other remaining combination of of 4 indexes.

\begin{figure}[!htb]
\centering
\vspace{-6pt}
\includegraphics[width=9cm]{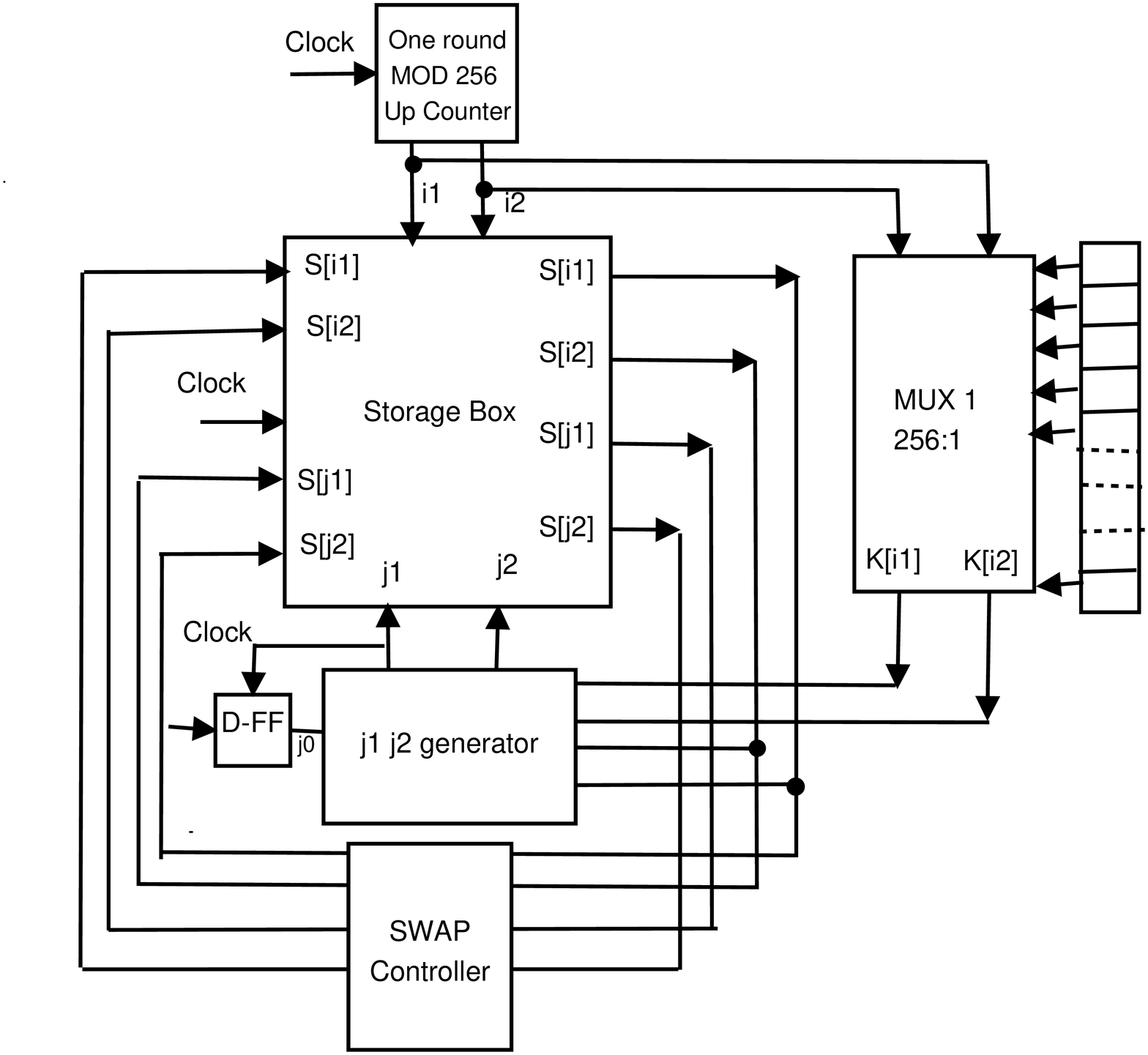}
\vspace{-6pt}
\caption{KSA unt of 2 byte per Clock Hardware}
\vspace{-8pt}
\label{fig:ksa2_fig}
\end{figure}

%-------------------- Table example ------------------------------
\begin{table}[]
\caption{Different cases for the data movement in the swap operation.} % title of Table
\centering  % used for centering table
\resizebox{9cm}{!}{%

    \begin{tabular}{|c|c|c|  }
        \hline
   $\#$    & Condition   & Register to register\\ 
~&~&Data movement\\\hline  
              
1	& $i_2 \neq j_1  \&  j_2 \neq i_1$ 	& $S_0[i_1]\rightarrow S_0[j_1], S_0[j_1]\rightarrow S_0[i_1],$\\
    ~     &$\& j_2\neq j_1$                        &$S_0[i_2] \rightarrow S_0[j_2], S_0[j_2]\rightarrow S0[i_2]$\\\hline
2	& $i_2 \neq j_1  \&  j_2 \neq i_1$ 	& $S_0[i_1]\rightarrow S_0[i_2]$,\\
~&$\& j_2= j_1$ &$ S_0[i_2]\rightarrow S_0[j_1] = S_0[j_2],$\\
~&~& $S_0[j_1]\rightarrow S_0[i_1]$\\\hline
3	&$i_2 \neq j_1  \&  j_2 = i_1$	& $S_0[i_1]\rightarrow S_0[j_1]$,\\
~&$ \& j_2\neq j_1$&$ S_0[i_2] \rightarrow S_0[i_1]=S_0[j_2],$\\
 ~&~& $S_0[j_1]\rightarrow S_0[i_2]$\\\hline
4	& $i_2 \neq j_1  \&  j_2=i_1$ 	&$S_0[i_1]\rightarrow S_0[i_2],$\\ 
~&$\& j_2= j_1$ &$S_0[i_2]\rightarrow S_0[i_1]=S_0[j_1]=S_0[j_2]$\\\hline
5	& $i_2=j_1  \&  j_2 \neq i_1$ 	&$S_0[i_1]\rightarrow S_0[j_2],$\\
~&$\& j_2\neq j_1$& $S_0[j_2]\rightarrow S_0[j_1]=S_0[i_2],$\\
~&~& $S_0[j_1]\rightarrow S_0[i_1]$\\\hline
6	& $i_2=j_1  \&  j_2 \neq i_1$	&$S_0[i_1]\rightarrow S_0[j_1]=S_0[i_2] =S_0[j_2],$\\
~&$ \& j_2=j_1$&$ S_0[j_1]\rightarrow S_0[i_1]$\\\hline
7	& $i_2=j_1  \&  j_2=i_1$	& No data Movement\\
~& $\& j_2\neq j_1$&~\\\hline
8	& $i_2=j_1  \&  j_2=i_1$ &Impossible\\
~& $ \& j_2=j_1$&~\\\hline

  \hline
       
    \end{tabular}
}
\label{table:swap_table} % is used to refer this table in the text
\end{table}
%--------------------- end of the example ------------------------
Following the data transfer of \ref{table:swap_table} we have designed a hardware behavioral model of swap controlling unit which has 4 input port to receive $S[i_1], S[i_2], S[j_1]$ and $S[j_2]$ from MUX and has 4 output to fed the swapped data to DEMUX unit. The Pictorial presentation is depicted on figure \ref{fig:ksa2_fig}.

\subsection{PRGA unit of 2 byte per clock architecture}
\begin{figure}[]
\centering
\vspace{-6pt}
\includegraphics[width=9cm]{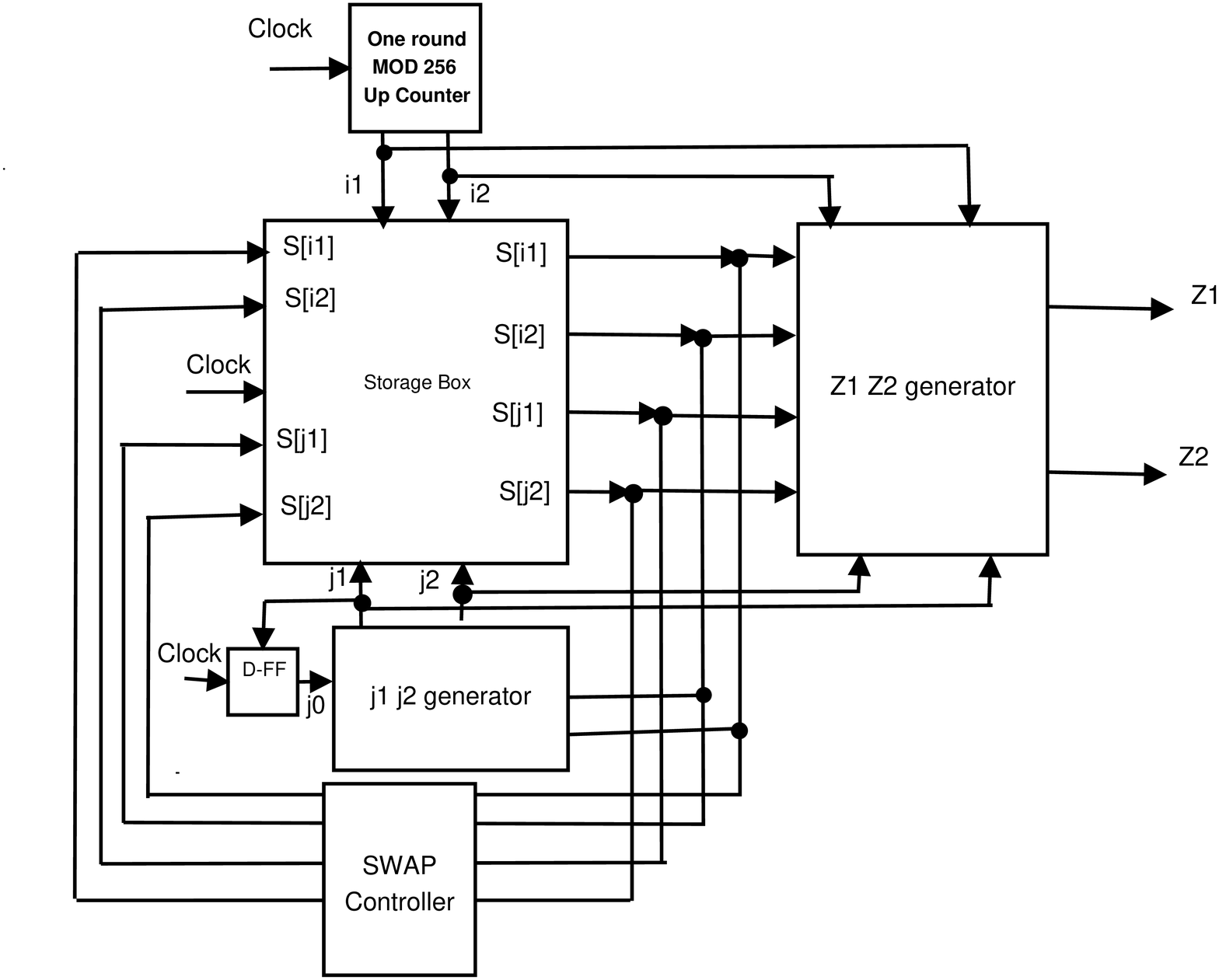}
\vspace{-6pt}
\caption{PRGA unit of 2 byte per Clock Hardware}
\vspace{-8pt}
\label{fig:prga2_fig}
\end{figure}

PRGA unit of 2 byte per clock architecture is very much smiler with the PRGA unit of 1 byte per clock architecture. Figure \ref{fig:prga2_fig} shows a schematic diagram of  the  design of the PRGA  unit. The Counter circuit generates $i_1$ and $i_2$ continuously and addresses $S[i_1]$ and $S[i_2]$ at at the same instance. The $j_1$ and $j_2$ is updated by the respective  $i_1$ and $i_2$ via  $j_1$ and $j_2$ generator and latches  the $S[j_1]$ and $S[j_2]$.  After generating $S[i_1], S[i_2], S[j_1]$ and $S[j_2]$, all of these signals has been fed to Swap Controlling Block. This Swap Block has decided which signal will be transfer to which address of S-box through the DEMUX circuit.
\subsubsection{ ~~~~$j_1$ and $j_2$ Generation of PRGA unit}
\begin{figure}[!htb]
\centering
\vspace{-6pt}
\includegraphics[width=9cm, height=5cm]{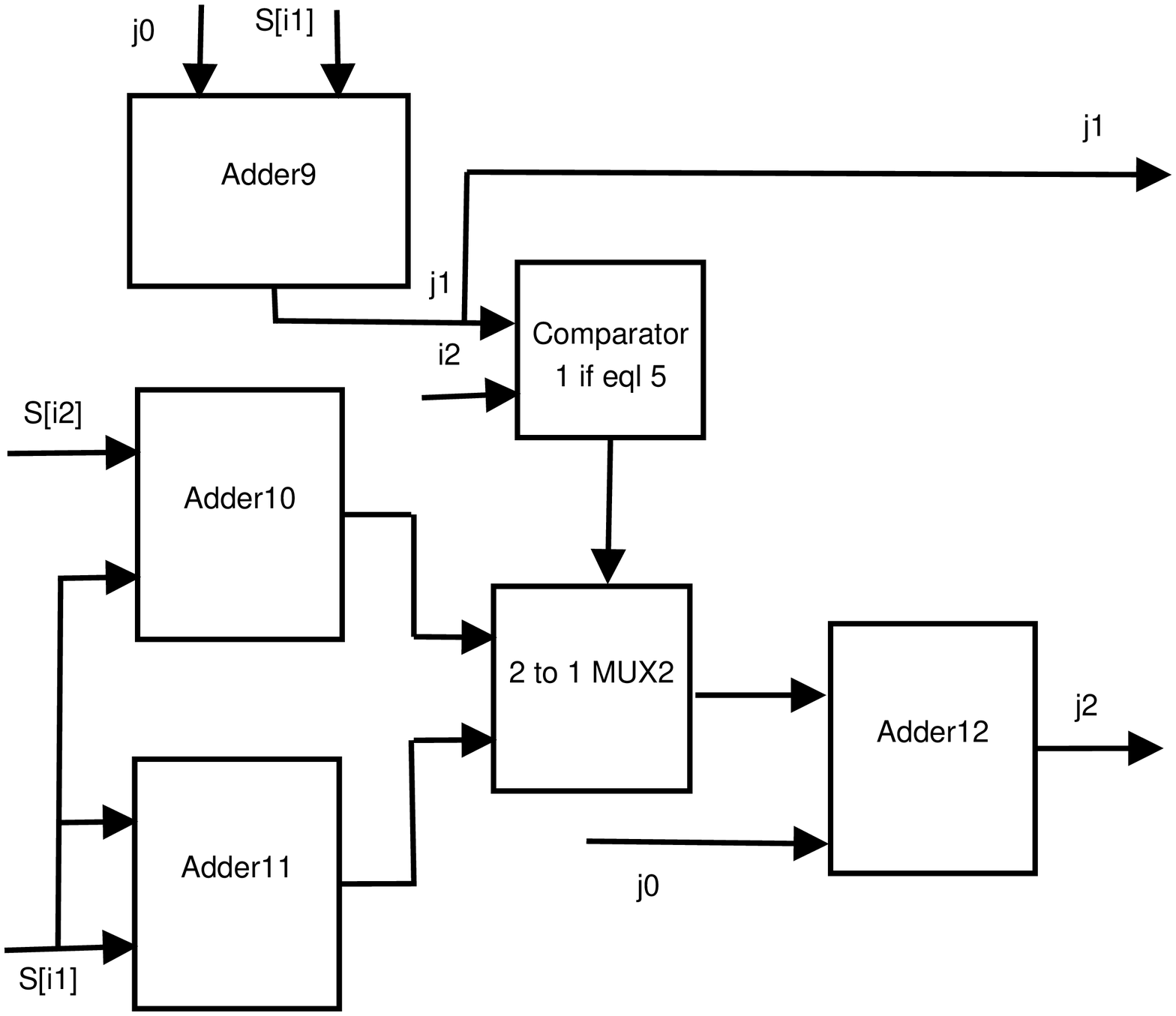}
\vspace{-6pt}
\caption{j1 and j2 generator of PRGA unit}
\vspace{-8pt}
\label{fig:j1_j2_prga_fig}
\end{figure}

This is a very simple circuit built by 4 adder blocks, 1 comparator and 1 MUX(2:1). As we seen in figure \ref{fig:j1_j2_prga_fig} $j_1$ computed very easily by Adder9 circuit but for the  $j_2$ computation we again need to see 2nd column step2 of table \ref{table:2steps_table} where the $j_2$ computation may be divided into the following two cases\\
\begin{equation}
\resizebox{.9\hsize}{!}{ $j_2=j_0+S_{0}[i_1]+S_{1}[i_2]\left\{\begin{array}{ll}\mbox{$ j_0+S_{0}[i_1]+S_{0}[i_2]$} & \mbox{if $i_2 \neq j_1$};\\
\mbox{$ j_0+S_{0}[i_1]+S_{0}[i_1]$} &  \mbox{$i_2 =j_1$}.\\
\end{array}\right. $}
\end{equation}
These two possible values from the two cases has been passed to MUX2 circuit through Adder10 and Adder11. The selecting input of the MUX2 is connected with Comparator5. The $j_1$ and $i_2$ have connected with the input of Comparator5. The Comparator5 takes the decision which value of $j_2$ will be passed to Adder12. Adder12 is computing the final $j_2$ by adding $j_0$ with it.

\subsubsection{~~~~~Swap Controlling block}
The Swap controlling circuit of PRGA process is identical with the swap circuit of KSA, as described in section 4.2.2.
\subsubsection{~~~~~$Z_1 \& Z_2$ Generator block}
\begin{figure}[!htb]
\centering
\vspace{-6pt}
\includegraphics[width=9cm, height=5cm]{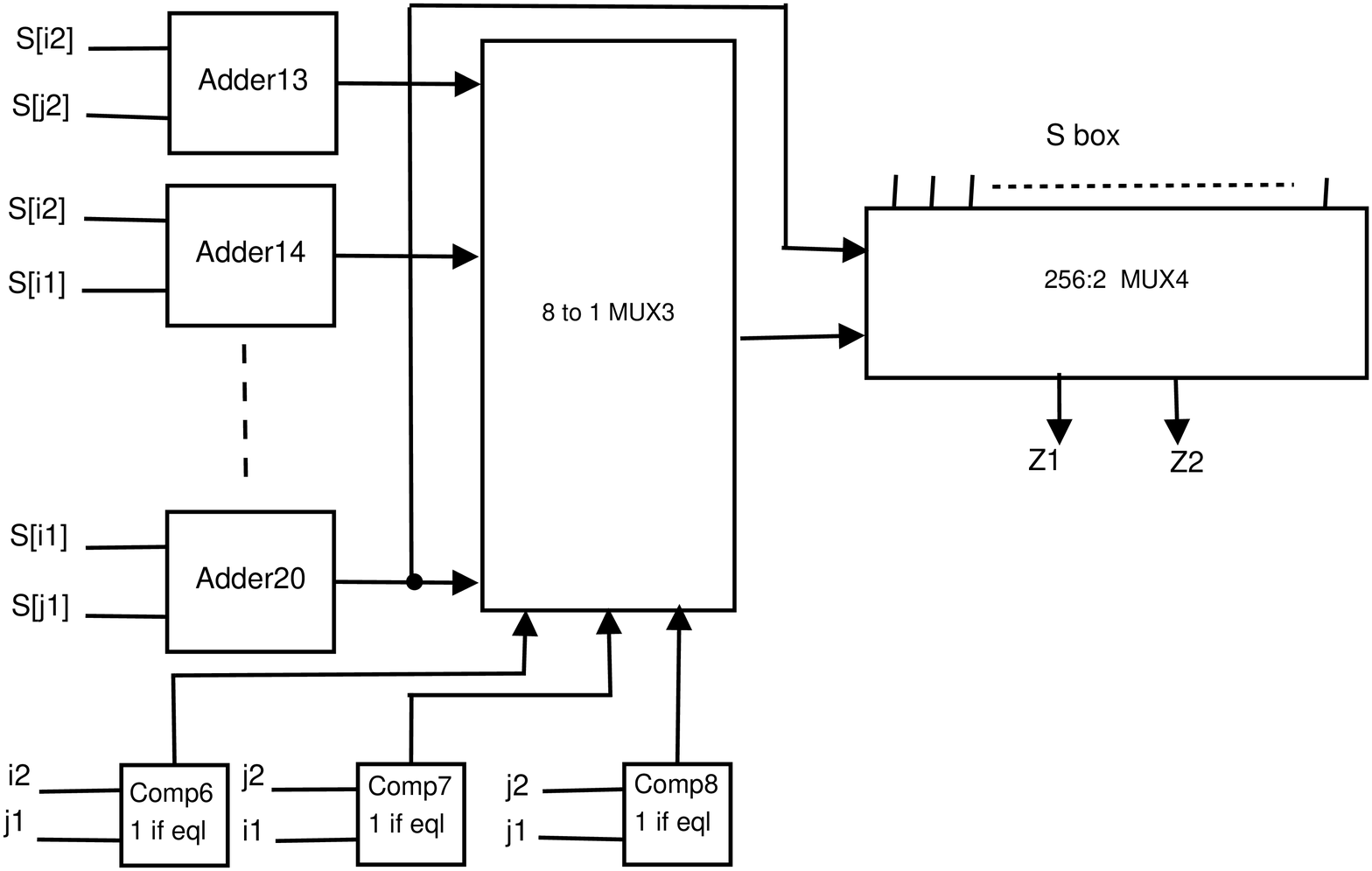}
\vspace{-6pt}
\caption{Z1 and Z2 generator of PRGA unit}
\vspace{-8pt}
\label{fig:z1_z2_fig}
\end{figure}

In step 4 of Table \ref{table:2steps_table}, we got 
\begin{equation}
Z_1=S_1[i_1] + S_1[j_1].
\end{equation}
As $S_0$ and $S_1$ has been differentiated by a swap process, the value of $Z_1$ can be evaluated like,\\
\begin{equation}
Z_1=S_0[j_1] + S_0[i_1]. 
\end{equation}
Thus the $Z_1$ can be computed by Adder20 of figure \ref{fig:z1_z2_fig} by adding $S_0[j_1]$ and $S_0[i_1]$. The output of Adder20 is connected with a 256:2 MUX2 (Merge of a two 256:1 MUX) which is used to retrieve the appropriate data from s-box. \\
Now the $Z_2$ computation is involved  as
\begin{equation}
Z_2 = S_2 [S_2[i_2] + S_2[j_2]] = S_2 [S_1[j_2] + S_1[i_2]] .
\end{equation}
As we used loop unrolling method, we need to jump directly from $S_0$ state to $S_2$ state. Here while we will compute $Z_2$, a swap of the 1st loop already has been computed so this important issue should be kept in our mind during the $Z_2$ computation. Where as the 2nd swap process could not make effect on $Z_2$ according to equation 5. So Considering all possible cases of $i_1, i_2, j_1$ and $j_2$ we are trying the compute the value of $S_1[i_2]$ and $S_1[j_2]$ in terms of $S_0$ state. The table \ref{table:z2_table} is showing all the details.

%-------------------- Table example ------------------------------
\begin{table}[]
\caption{Different cases for the data movement for the $Z_2$ computation.} % title of Table
\centering  % used for centering table
\resizebox{9cm}{!}{%

    \begin{tabular}{|c|c|c|  }
        \hline
   $\#$    & Condition   & Register to register\\ 
~&~&Data movement\\\hline  
              
1	& $i_2 \neq j_1  \&  j_2 \neq i_1$ 	& $S_1[i_2] = S_0[i_2]$,\\
 ~&$j_2\neq j_1$&$S_1[j_2] = S_0[j_2]$\\\hline
2	& $i_2 \neq j_1  \&  j_2 \neq i_1$ 	& $S_1[i_2] = S_0[i_2]$,\\
~&$\& j_2= j_1$ &$ S_1[j_2] = S_0[i_1]$\\\hline
3	&$i_2 \neq j_1  \&  j_2 = i_1$	& $S_1[i_2] = S_0[i_2],$\\
~&$ \& j_2\neq j_1$&$S_1[j_2] = S_0[j_1]$\\\hline
4	& $i_2 \neq j_1  \&  j_2=i_1$ 	&$S_1[i_2] = S_0[i_2],$\\ 
~&$\& j_2= j_1$ &$S_1[j_2]=S_0[j_1]$\\\hline
5	& $i_2=j_1  \&  j_2 \neq i_1$ 	&$S_1[i_2]=S_0[i_1],$\\
~&$\& j_2\neq j_1$& $S_1[j_2] = S_0[j_2]$\\\hline
6	& $i_2=j_1  \&  j_2 \neq i_1$	&$S_1[i_2]=S_0[i_1],$\\
~&$ \& j_2=j_1$&$S_1[j_2] = S_0[i_1]$\\\hline
7	& $i_2=j_1  \&  j_2=i_1$	& $S_1[i_2] = S_0[i_1]$,\\
~& $\& j_2\neq j_1$&$S_1[j_2]=S_0[j_1]$\\\hline
       
    \end{tabular}
}
\label{table:z2_table} % is used to refer this table in the text
\end{table}
%--------------------- end of the example ------------------------
The circuit of the $Z_2$ computation can be realized using a 8:1 MUX (named as MUX3). 3 comparator circuit is connected with the 3 selecting input of MUX3. Comparator 6,7 and 8 comparing (i) $i_2$ and $j_1$, (ii) $j_2$ and $i_1$, (iii) $j_2$ and $j_1$. The output of MUX3 is connected with the selecting input of MUX4 to address the proper appropriate element of s-box as $Z_2$ key. The hardware description of $Z_2$ computation is shown in figure \ref{fig:z1_z2_fig}.

\subsection{Timing Analysis of PRGA of 2 byte per clock design}
Again a MOD 256 up counter shown in figure \ref{fig:prga2_fig} is designed which is very much identical with the up counter \ref{fig:prga} but instead of single output it has two consecutive outputs named as $i_1$ and $i_2$. That $i_1$ starts from 1, goes up to 255 skipping all the even value between the said range and it repeats again until the plaintext sequence has been stopped. The $i_2$ starts from 2, goes up to 254  skipping all the odd value between the said range and then it repeats again from 0 to 254 skipping the same odd numbers. This means the said counter generates two consecutive counting values at the same clock instance. The details timing analysis has been shown below. \\
\begin{itemize}
\item Rising edge of  $\phi_0$: Initialize $j_0=0$ and $i_0$=0.\\
\item Falling edge of $\phi_0$: Start counter $i_1$=1 and $i_2$=2.\\
\item Rising edge of  $\phi_1$: $j_1$=$(j_0 + S_0[i_1])$ \%\ 256; $j_2=j_0+S_0[i_1]+S_1[i_2]$, temp=$i_1$, temp\_next=$i_2$.\\
\item Falling edge of $\phi_1$: $i_3=$3; and $i_4=4$, Swap Occurred;\\
\item Rising edge of  $\phi_2: j_3=(j_2+S_2[i_3]) \%\ 256;$ $j_4=(j_2+S_2[i_3]+S_3[i_4]) \%256,$ temp=$i_3$, temp\_next=$i_4$, $Z_1=S_1(S_0[i_1]+S_0[j_1])$ \%256, $Z_2 = S_2[S_1[i_2] + S_1[j_2]]$\%\ 256.\\
\item Falling edge of $\phi_2$: $i_5=$5; and $i_6=6$, Swap Occurred; \\
\item Rising edge of  $\phi_3: j_5=(j_4+S_4[i_5]) \%\ 256;$ $j_6=(j_4+S_2[i_5]+S_5[i_6]) \%256,$ temp=$i_5$, temp\_next=$i_6$, $Z_3=(S_2[i_3]+S_2[j_3])$\%\ 256, $Z_4=S_3[S_2[i_4]+S_2[j_4]$\%\ 256.\\
\end{itemize}

%\begin{figure}[!htb]
%\centering
%\vspace{-6pt}
%\includegraphics[width=9cm,  height=3cm]{./figure/timing2.png}
%\vspace{-6pt}
%\caption{PRGA unit of 2 byte per Clock Hardware}
%\vspace{-8pt}
%\label{fig:timing2_fig}
%\end{figure}

\begin{tikztimingtable}[
    timing/slope=0,         % no slope
    timing/coldist=.0pt,     % column distance
    xscale=2.99,yscale=3.1, % scale diagrams
    semithick               % set line width
  ]
  \scriptsize  & 8{C}                              \\
  % $\overline{\mbox{Q}}$ &  LHLLLHH                  \\
\extracode
 %\makeatletter
 \begin{pgfonlayer}{background}
\begin{scope}[gray,semitransparent,semithick]
   
    \foreach \x in {1,...,8}
      \draw (\x,1) -- (\x,-2.8);
    % similar: \vertlines{1,...,6}
  \end{scope}
    \node [anchor=south east,inner sep=0pt]
    at (0.9,0.2) {CLK};

  \node [anchor=south east,inner sep=0pt]
    at (1.3,-1.45) {\tiny Initialize};
  \node [anchor=south east,inner sep=0pt]
    at (1.3,-1.70) {\tiny $i_0=0$};
  \node [anchor=south east,inner sep=0pt]
    at (1.3,-1.95) {\tiny $j_0=0$};

  \node [anchor=south east,inner sep=0pt]
    at (2.3,-0.45) {\tiny Start};
  \node [anchor=south east,inner sep=0pt]
    at (2.3,-0.70) {\tiny Counter};
  \node [anchor=south east,inner sep=0pt]
    at (2.3,-0.95) {\tiny $i_1=1$};
  \node [anchor=south east,inner sep=0pt]
    at (2.3,-1.20) {\tiny $i_2=2$};

  \node [anchor=south east,inner sep=0pt]
    at (3.6,-1.65) {\tiny $j_1=(j_0+$};
  \node [anchor=south east,inner sep=0pt]
    at (3.6,-1.90) {\tiny $S_0[i_1]$};
  \node [anchor=south east,inner sep=0pt]
    at (3.6,-2.15) {\tiny $\%256$};
  \node [anchor=south east,inner sep=0pt]
    at (3.6,-2.40) {\tiny $j_2=(j_1+$};
  \node [anchor=south east,inner sep=0pt]
    at (3.6,-2.65) {\tiny $S_1[i_2]$};
  \node [anchor=south east,inner sep=0pt]
    at (3.6,-2.90) {\tiny $\%256$};

  \node [anchor=south east,inner sep=0pt]
    at (4.3,-0.45) {\tiny $i_3=3$};
  \node [anchor=south east,inner sep=0pt]
    at (4.3,-0.70) {\tiny $i_4=4$};
  \node [anchor=south east,inner sep=0pt]
    at (4.3,-1.00) {\tiny Swap};
  \node [anchor=south east,inner sep=0pt]
    at (4.3,-1.25) {\tiny $Z_1, Z_2$};
  \node [anchor=south east,inner sep=0pt]
    at (4.3,-1.50) {\tiny executed};

  \node [anchor=south east,inner sep=0pt]
    at (5.6,-1.65) {\tiny $j_3=(j_2+$};
  \node [anchor=south east,inner sep=0pt]
    at (5.6,-1.90) {\tiny $S_2[i_3]$};
  \node [anchor=south east,inner sep=0pt]
    at (5.6,-2.15) {\tiny $\%256$};
  \node [anchor=south east,inner sep=0pt]
    at (5.6,-2.40) {\tiny $j_4=(j_3+$};
  \node [anchor=south east,inner sep=0pt]
    at (5.6,-2.65) {\tiny $S_3[i_4]$};
  \node [anchor=south east,inner sep=0pt]
    at (5.6,-2.90) {\tiny $\%256$};

  \node [anchor=south east,inner sep=0pt]
    at (6.3,-0.45) {\tiny $i_5=5$};
  \node [anchor=south east,inner sep=0pt]
    at (6.3,-0.70) {\tiny $i_6=6$};
  \node [anchor=south east,inner sep=0pt]
    at (6.3,-1.00) {\tiny Swap};
  \node [anchor=south east,inner sep=0pt]
    at (6.3,-1.25) {\tiny $Z_3, Z_4$};
  \node [anchor=south east,inner sep=0pt]
    at (6.3,-1.50) {\tiny executed};

  \node [anchor=south east,inner sep=0pt]
    at (7.6,-1.65) {\tiny $j_5=(j_4+$};
  \node [anchor=south east,inner sep=0pt]
    at (7.6,-1.90) {\tiny $S_4[i_5]$};
  \node [anchor=south east,inner sep=0pt]
    at (7.6,-2.15) {\tiny $\%256$};
  \node [anchor=south east,inner sep=0pt]
    at (7.6,-2.40) {\tiny $j_6=(j_5+$};
  \node [anchor=south east,inner sep=0pt]
    at (7.6,-2.65) {\tiny $S_3[i_6]$};
  \node [anchor=south east,inner sep=0pt]
    at (7.6,-2.90) {\tiny $\%256$};

  \node [anchor=south east,inner sep=0pt]
    at (8.3,-0.45) {\tiny $i_7=7$};
  \node [anchor=south east,inner sep=0pt]
    at (8.3,-0.70) {\tiny $i_8=8$};
  \node [anchor=south east,inner sep=0pt]
    at (8.3,-1.00) {\tiny Swap};
  \node [anchor=south east,inner sep=0pt]
    at (8.3,-1.25) {\tiny $Z_5, Z_6$};
  \node [anchor=south east,inner sep=0pt]
    at (8.3,-1.50) {\tiny executed};

 \end{pgfonlayer}
\label{fig:timing2_fig}
\end{tikztimingtable}%

\subsection{Dynamic KSA-PRGA Architecture (Design 4)}
Taking the advantage of selfsame architecture of KSA and PRGA process, the same approach has been adopted in the 2 byte per clock architecture. The KSA process in transformed into PRGA process taking the acknowledgment of $'prga~enable' (prga\_en)$ signal. Only the counter of said design is slightly different from 1 byte dynamic KSA-PRGA architecture of section 3.5. Two outputs of the said counter $i_1$ and $i_2$ count two consecutive counts starts from count $'0'$ and $'1'$ value respectively following an initialization clock of KSA process.  At the 129th clock $i_1$ and $i_2$ reaches at count $254$ and $255$ respectively. At the 130th clocks $i_1$, $i_2$ and $j_1$, $j_2$ are initialized by $0$ and $prga~enable$ goes high for the successive PRGA process. Since swap is not required for this initialization state, swap process has been stopped for this single clock duration.  After this initialization again the $i_1$ and $i_2$ counts has been started with $'1'$ and $'2'$ counts. This consecutive counts is going on while we need key streams form PRGA process. The timing diagram is shown in \ref{fig:timing2_fig}
\begin{figure}[!htb]
\centering
\vspace{-6pt}
\includegraphics[width=9cm,  height=10cm]{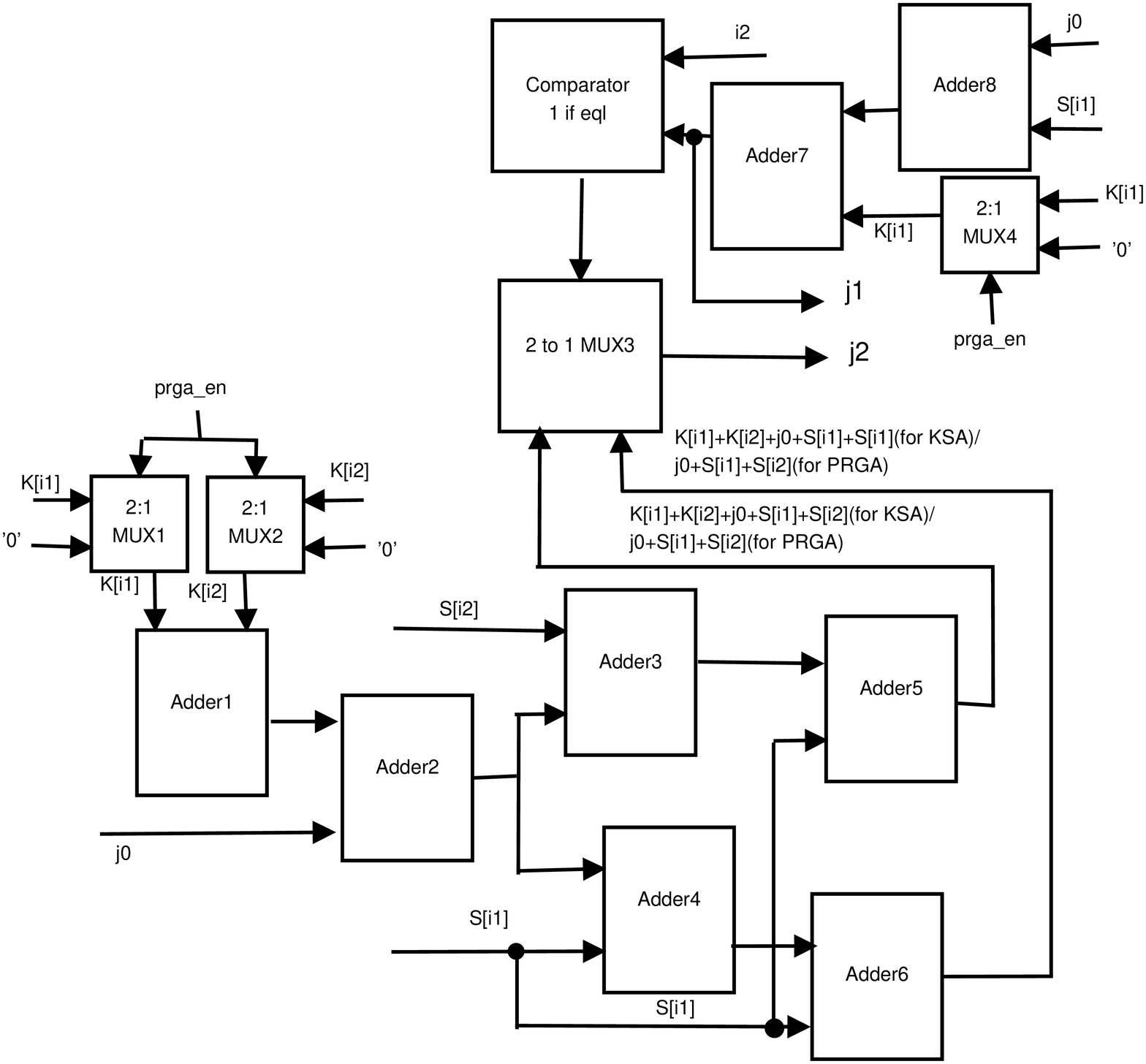}
\vspace{-6pt}
\caption{$j_1$ and $j_2$ Generator of Dynamic KSA-PRGA of 2 byte per Clock Hardware}
\vspace{-8pt}
\label{fig:dkp_j1_j2}
\end{figure}
\subsubsection{~~~~~Hardware overview}
\begin{figure}[!htb]
\centering
\vspace{-6pt}
\includegraphics[width=9cm,  height=8cm]{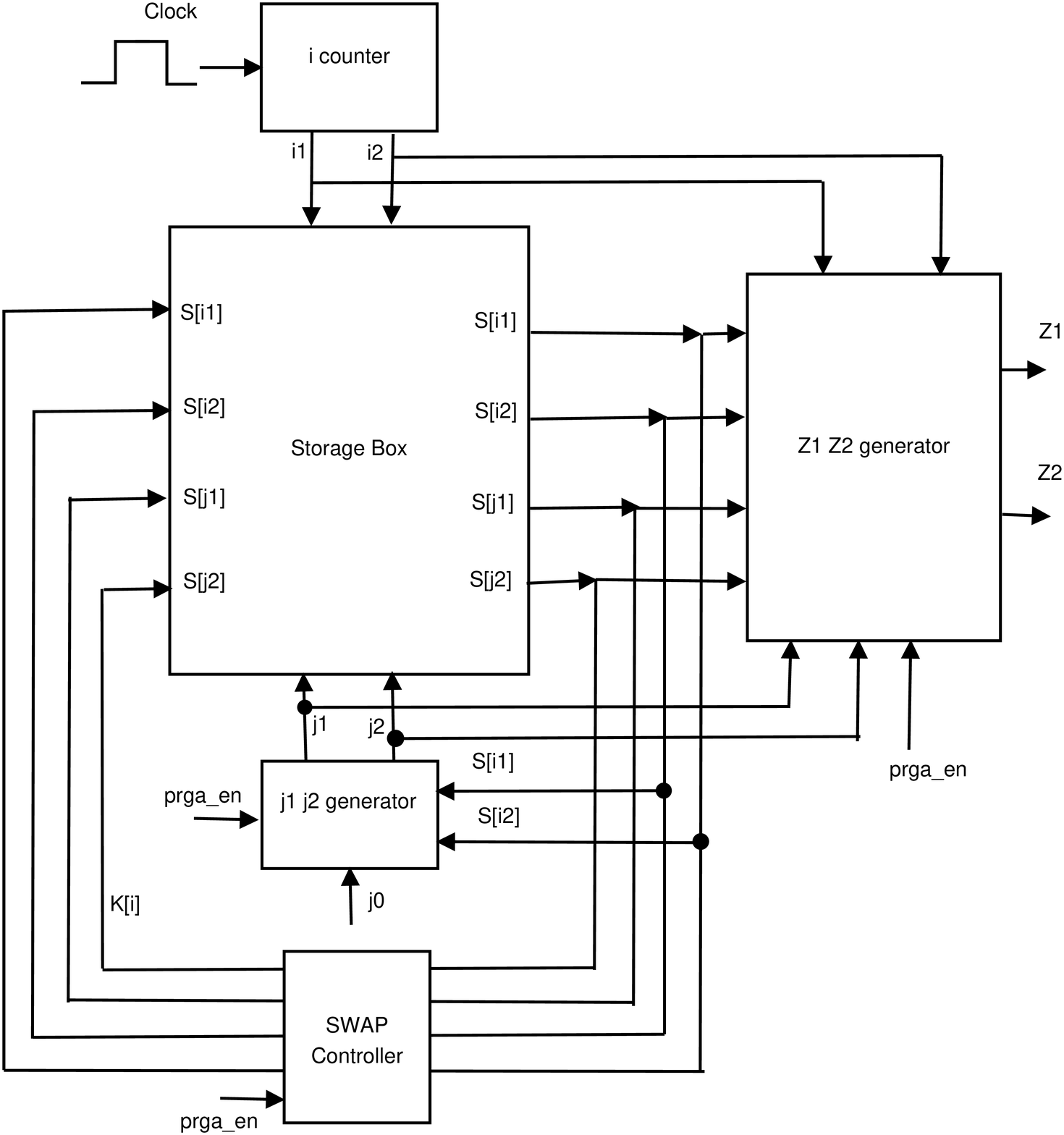}
\vspace{-6pt}
\caption{Dynamic KSA-PRGA of 2 byte per Clock Hardware}
\vspace{-8pt}
\label{fig:dynamic2}
\end{figure}

The hardware of $j_1$ and $j_2$ generator is shown in figure \ref{fig:dkp_j1_j2}. 
Bu MUX1 and MUX2 the $prga~enable$ signal decides whether $K[i_1]+K[i_2]$ or $'0'$ will be passed to Adder1. If $prga~enable$ goes low then $K[i_1]+K[i_2]$ will be added with signal $j_0+S[i_1]+S[i_1]$ (for $i_2=j_1$) and with signal $j_0+S[i_1]+S[i_2]$ (for $i_2 \neq j_1$). Otherwise $'0'$ will be added with the said signal for both of the mentioned conditions using the Adder2, Adder3, Adder4, Adder5, Adder6 blocks.Among these two values of $j_2$, which one will passes as a final $j_2$ output is decided by Comparator circuit which has two input $i_2$ and $j_1$. The $j_1$ computation is smiler with section 3.5 architecture figure \ref{fig:dynamic}. The top level hierarchy of Dynamic KSA-PRGA architecture is shown in figure \ref {fig:dynamic2}. The functionality and pictorial representation of Storage Block unit, $Z_1$ and $Z_2$ generator has been already discussed and shown in section 4.1 and figure \ref{fig:s_box2}, section 4.3.3figure \ref{fig:z1_z2_fig} respectively. The swap control unit also has been discussed in section 4.3.2.
\section{1 byte per clock architecture as 4 co-processor (Design 5)}
\label{design5}
\begin{figure}[!htb]
\centering
\vspace{-6pt}
\includegraphics[width=9cm,  height=10cm]{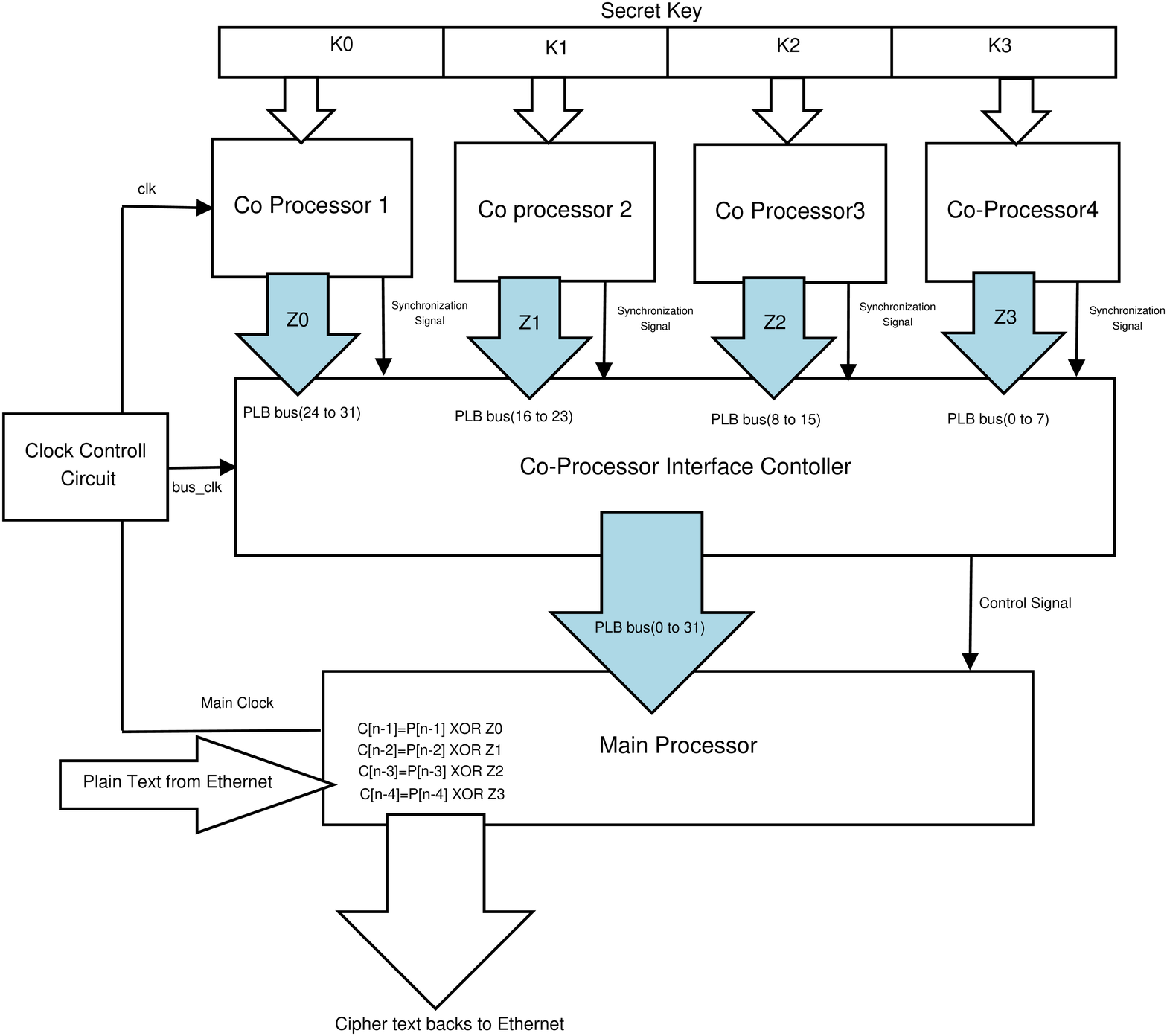}
\vspace{-6pt}
\caption{4 RC4 co-processor with main processor}
\vspace{-8pt}
\label{fig:4_co}
\end{figure}

The back ground of this approach is $"how~many~bytes~can~be~generated~in~a~single~clock?"$. The more loop unrolling can make the circuit more complex in terms of latency and critical path, so we thought if we can accommodate four RC4 co-processor communicating with main processor to en-cipher 4 plain text character coming from the ethernet (or from other interfaces) at a single clock. 4 co-processor environment has been chose as we have 32 bit main processor and 32 bus availability. Here at an instant only 4 Z can be communicated with main processor through bus. See the figure \ref{fig:4_co} where 4 separate RC4 Co-processors work breaking the secret keys into same sizes (different size can also be accommodated according to the requirement of user). The four separated RC4 blocks generate 4 keys (Z) at a single pulse. So we can say the architecture can generate 4 bytes at single clock. After computing of Zs, they are passed to the main processor to XOR those Zs (keys) with the plain text coming from ethernet. When the ciphering has been done, again those ciphers have been send to the unsecured ethernet environment.\\
\section{2 byte per clock architecture as 2 co-processor (Design 6)}
\label{design6}
\begin{figure}[!htb]
\centering
\vspace{-6pt}
\includegraphics[width=9cm,  height=10cm]{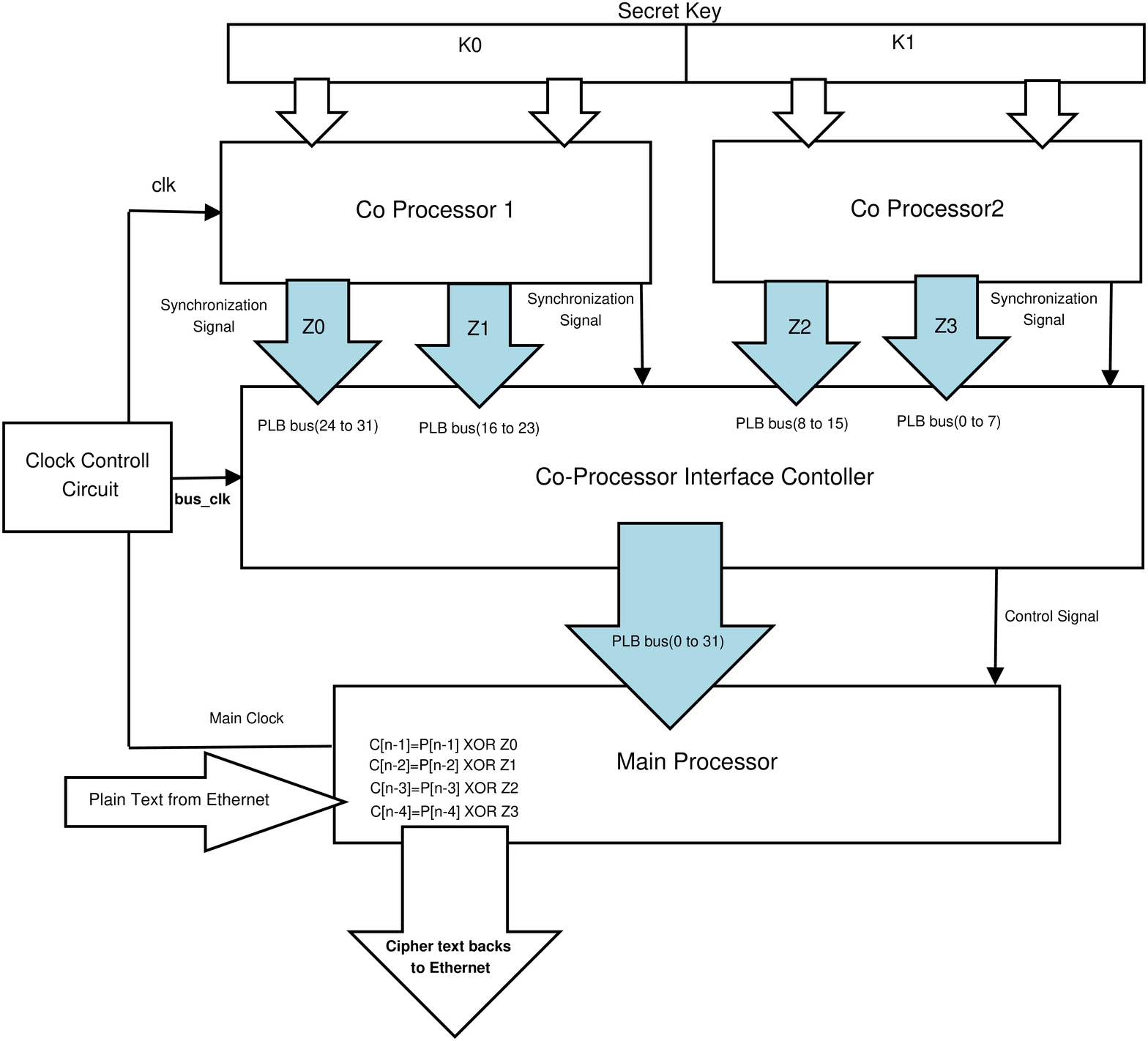}
\vspace{-6pt}
\caption{2 RC4 co-processor with main processor}

\vspace{-8pt}
\label{fig:2_co}
\end{figure}

The same 4 byte per clock throughput can be achieved, if we used two 2 byte per clock architecture instead of using four 1 byte per clock architecture. here in figure \ref{fig:2_co} two 2 byte per clock architecture can generate 4 keys at a single clock. The hardware usage and power consumption of proposed four different architecture is shown in table \ref{table:hw_table} and \ref{table:pw_table}.

\section{Critical Path Analysis}
Here the critical path analysis
%---------------------------------------------------------------------------------
\begin{table}[]
\caption{Critical path analysis} % title of Table
\centering  % used for centering table
\resizebox{9cm}{!}{%

    \begin{tabular}{|c|c|c|  }
        \hline
Architectures& 1 byte  & 2 byte \\
~&per Clock&per Clock\\\hline

Critical Path & 4.127&5.15\\
(ns) &~&~\\
\hline
    \end{tabular}
}
\label{table:critical_path} % is used to refer this table in the text
\end{table}
%--------------------- end of the example ------------------------
\section{Results and implementation}
\label{ram}
The four kind of architetcure have been proposed, such as, 1)1 byte per clock, 2)1 byte per clock Dynamic KSA-PRGA(DKP) clock, 3)2 byte per clock. and finally 4)2 byte per clock Dynamic KSA-PRGA(DKP) clock. The hardware usage and power consumption of these proposed four different architectures are shown in table \ref{table:hw_table} and \ref{table:pw_table}. All of the architecture has been imported in parallel processing environment where main processor handling the data interface part and co-processor are taking care of RC4 algorithm part. We have seen the trade off between throughput, power consumption and throughput, resource usage. We have also observed how the power consumption and resource usage increases with the increasing of the number of co-processor in tables  \ref{table:diff_hw_table}, \ref{table:diff_pw_table}, \ref{table:diff2_hw_table}, and \ref{table:diff2_pw_table}.

%----------------------------------------------------------------------------------
\begin{table}[!htb]
\caption{Power consumption of Proposed Architectures} % title of Table
\centering  % used for centering table
\resizebox{9cm}{!}{%

    \begin{tabular}{|c|c|c|c|   }
        \hline
Milli &	Static&	Dynamic&	Total\\
watt&	 Power&	Power&	 Power\\\hline
Design 1&	970.8&	206.4&	1177.2\\\hline
design 2&	914.87&	79.85&	994.72\\\hline
Design 3&	976.34&	446.38&	1422.71\\\hline
Design 4&	 953.94&	109.91&	1063.85\\\hline
Design 5&	 1004.74&	159.92&	1164.66\\\hline
Design 6&	 1016.96&	194.61&	1211.58\\

\hline
    \end{tabular}
}
\label{table:pw_table} % is used to refer this table in the text
\end{table}
%--------------------- end of the example ------------------------

   \begin{tikzpicture}
        \begin{axis}[
            symbolic x coords={a small bar, a medium bar, a large bar},
            xtick=data
          ]
            \addplot[ybar,fill=blue] coordinates {
                (a small bar,   42)
                (a medium bar,  50)
                (a large bar,   80)
            };
        \end{axis}
    \end{tikzpicture}

%----------------------------------------------------------------------------------
\begin{table}[!htb]
\caption{Hardware usage of Proposed Architectures} % title of Table
\centering  % used for centering table
\resizebox{9cm}{!}{%

    \begin{tabular}{|c|c|c|c|   }
        \hline
\# &	Slice&	LUTs&	Fully used \\
~&~&~&LUTs FF pairs\\\hline
Design 1&	4173&	14588&	4149\\\hline
Design 2&	2094	&5383	&2078\\\hline
Design 3&	4178&	33115&	4168\\\hline
Design 4&	2123&	15452&	2104\\\hline
Design 5&	17964&	49545&	49886\\\hline
Design 6&	5682&	36958&	37420\\
\hline
    \end{tabular}
}
\label{table:hw_table} % is used to refer this table in the text
\end{table}
%--------------------- end of the example ------------------------

%----------------------------------------------------------------------------------
\begin{table}[!htb]
\caption{Hardware usage of 1 byte per clock architecture for different number of co-processor} % title of Table
\centering  % used for centering table
\resizebox{9cm}{!}{%

    \begin{tabular}{|c|c|c|c|c|c|  }
        \hline
Milli &	Only &	1 Co-&	2 Co-&	3 Co-&	4 Co-\\
watt&	Algorithm &	Processor&	-Processor&	-Processor&	-Processor\\
~&	Core&	~&	~&	~&	Design 5\\\hline
Static &	914.87&	1000.45&	1002.35&	1004.12&	1004.74\\
 Power&	~&	~&	~&	~&	~\\\hline
Dynamic &	052.86&	83.25&	117.39&	148.95&	159.92\\
 Power&	~&	~&	~&	~&	~\\\hline
Total &	994.72&	1083.69&	1119.74&	1153.07&	1164.66\\
 Power&	~&	~&	~&	~&	~\\\hline

    \end{tabular}
}
\label{table:diff_hw_table} % is used to refer this table in the text
\end{table}
%--------------------- end of the example ------------------------

%----------------------------------------------------------------------------------
\begin{table}[!htb]
\caption{Power consumption of 1 byte per clock architecture for different number of co-processor} % title of Table
\centering  % used for centering table
\resizebox{9cm}{!}{%

    \begin{tabular}{|c|c|c|c|c|c|  }
        \hline
Milli &	Only &	1 Co-&	2 Co-&	3 Co-&	4 Co-\\
watt&	Algorithm &	Processor&	-Processor&	-Processor&	-Processor\\
~&	Core&	~&	~&	~&	Design 5\\\hline
Slice &	4139&	5602&	9740&	13842&	17964\\
Registers&   	~&	~&	~&	~&	~\\\hline
Slice&	12560&	13986&	26013&	38217&	49545\\
 LUTS&   	~&	~&	~&	~&	~\\\hline
Slice LUT-&   	4132&	14510&	26396&	38782&	49886\\
F/ F pairs&   	~&	~&	~&	~&	~\\\hline

    \end{tabular}
}
\label{table:diff_pw_table} % is used to refer this table in the text
\end{table}
%--------------------- end of the example ------------------------

%----------------------------------------------------------------------------------
\begin{table}[!htb]
\caption{Hardware usage of 2 byte per clock architecture for different number of co-processor} % title of Table
\centering  % used for centering table
\resizebox{9cm}{!}{%

    \begin{tabular}{|c|c|c|c|  }
        \hline
Milli &	                    Only &	           1 Co-&	             2 Co-\\
watt&	                   Algorithm &	Processor&	-Processor\\
~&	                   Core&	           ~&	           Design 6\\\hline
Static Power&	       953.94&	           1011.457&	1016.96\\\hline
Dynamic Power&	109.91&	158.48&	194.61\\\hline
Total Power&	1063.85&	1173.05&	1211.58\\
\hline
    \end{tabular}
}
\label{table:diff2_hw_table} % is used to refer this table in the text
\end{table}
%--------------------- end of the example ------------------------

\begin{figure}[!htb]
\centering
\vspace{-6pt}
\includegraphics[width=9cm,  height=8cm]{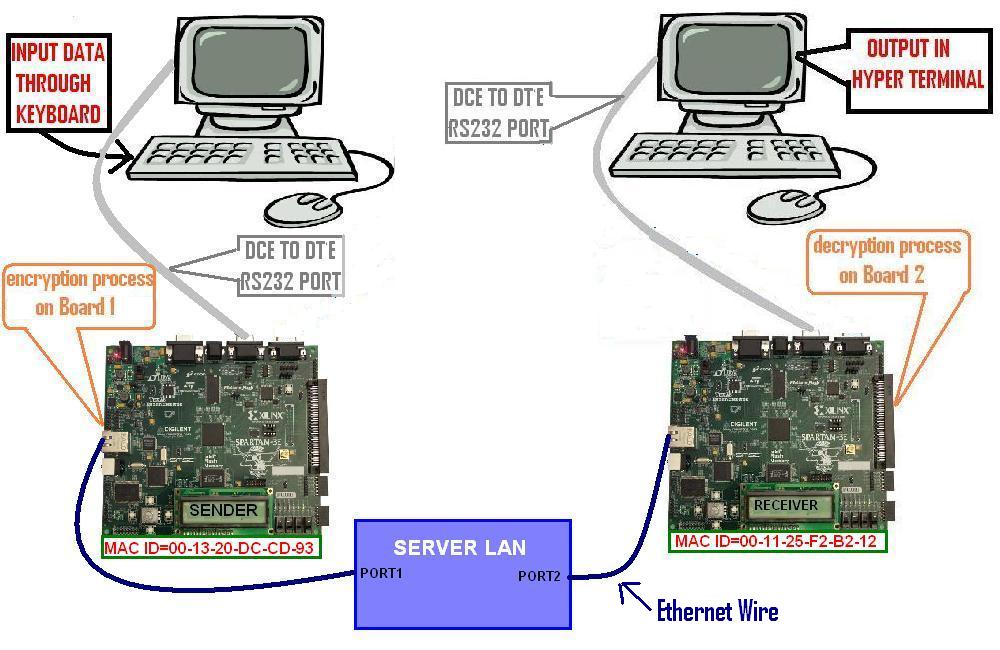}
\vspace{-6pt}
\caption{Experimental setup of FPGA based secured data communication}

\vspace{-8pt}
\label{fig:b_to_b}
\end{figure}

%----------------------------------------------------------------------------------
\begin{table}[!htb]
\caption{Power consumption of 2 byte per clock architecture for different number of co-processor} % title of Table
\centering  % used for centering table
\resizebox{9cm}{!}{%

    \begin{tabular}{|c|c|c|c|  }
        \hline
Milli &	                    Only &	           1 Co-&	             2 Co-\\
watt&	                   Algorithm &	Processor&	-Processor\\
~&	                   Core&	           ~&	           Design 6\\\hline
Slice Registers&	2123&	3386&	5682\\\hline
Slice LUTs&	15452&	16821&	36958\\\hline
Slice LUT-F/ F pairs&   	2104&	17420&	37420\\

\hline
    \end{tabular}
}
\label{table:diff2_pw_table} % is used to refer this table in the text
\end{table}
%--------------------- end of the example ------------------------
All the designs has been implemented on Virtex(ML505, lx110t) and Spartan 3E(XC3S500e) FPGA board using ISE 14.4 and EDK 14,4 tool. The power and  resource result is shown for Virtex board. The co processor has been coded by VHDL language and the main processor functionality has projected by system.c language. Two Xilinx FPGA Spartan3E (XC3S500e-FG320) boards, each with RC4 encryption and decryption engines separately and shown in Figure \ref{fig:b_to_b}, are connected through Ethernet ports and to respective hyper terminals (DTE) through RS 232 ports. 

\section{Comparison with Existing Designs}
In table \ref{table:compare} we have compared the throughput of our design with existing work till date.  In section \ref{design12} we have seen a 1 byte per clock architecture where KSA is executing within 257 clock after an initial lag of 1 clock. The PRGA is computing 1 byte after a lag of 2 initial clock.  The KSA process in section \ref{design34} named as of 2 byte per clock architecture can iterate 256 loops within 128 clock after an initial clock delay and the PRGA of that said architecture can execute 2 bytes in single clock after a lag of 2 initial clocks. From the table \ref{table:compare} we can see Ref  \cite{IEEE:b}, \cite{patent:matthews} are 3 byte per clock design. Article \cite{dp:math}, \cite{springerlink:one_byte} and design 1, design 2 have same throughput of 1 byte per clock but more sophisticatedly  article \cite{springerlink:one_byte} and design1, design 2 have slightly better throughput as mention in the said table \ref{table:compare} where as article \cite{ieee:two_byte}  and design 3, design 4 have same throughput of 2 byte per clock. More specifically our proposed design 3 and design 4 has slightly better throughput than \cite{ieee:two_byte}.

\begin{table*}[ht]
\caption{Comparison of various performance metrics with existing designs} % title of Table
\centering  % used for centering table
\resizebox{16cm}{!}{%

    \begin{tabular}{|c|c|c|c|c|c|c| }
        \hline
     Features   & \multicolumn{6}{c|}{Number of Clock Cycles}\\  
\cline{2-7}
 ~  & Ref  \cite{IEEE:b}, \cite{patent:matthews}  &Ref  \cite{dp:math} &    Ref. \cite{springerlink:one_byte}      &          \cite{ieee:two_byte}                                        & Design 1                                   & Design 3  \\ 
 ~  &~   &~&~                                                             &~                                  & \& Design 2                                &  \& Design 4  \\ \hline  
              
KSA	                  &256X3=768	&3 + 256 = 259    & 1 + 256= 257  & 1+256  &1 + 256= 257                         & 1+128=129\\\hline
KSA 	      &3	                        &1 + 3/256 	      &1 + 1/256          &1+1/256             &1 + 1/256                               &1/2+ 1/256\\
per byte	      &~	                        &~ 	      &~               &~                     &~                                             &~\\\hline
PRGA for   & 3n	            &3 + n  	      &2 + n	& n/2+2                      &2 + n                                       & 2+n/2\\
N byte	      &~	                        &~ 	      &~        &~    &~                                                                                    &~\\\hline
PRGA     &3	                        &1 + 3/n	      &1 + 2/n	&1/2+2/n                       &1 + 2/n                                   &1/2+2/n\\
per byte	      &~	                        &~ 	      &~                     &~               &~                                             &~\\\hline
RC4 for     & 3n+768	            &259+(3+n)	      &257+(2+n)          &257+(2+n/2)           &257+(2+n)                              &129+(2+n/2)\\
N byte	      &~	                        &~ 	 &~     &~        &~                                                                                &~\\\hline
Per byte	    & 3+768/n	            &1 + 262/n	      &1 + 259/n  &~1/2 + 259/n        &1 + 259/n                               &1/2+131/n \\ 
output	      &~	                        &~ 	      &~      &~  &~                                                                   &~\\
from RC4	      &~	                        &~ 	      &~      &~  &~                                                                   &~\\
\hline
       
    \end{tabular}
}
\label{table:compare} % is used to refer this table in the text
\end{table*}
%--------------------- end of the example ------------------------

\section{Conclusion}
The conclusion goes here.

\appendices
\section{Proof of the First Zonklar Equation}
Appendix one text goes here.

\section{}
Appendix two text goes here.

\ifCLASSOPTIONcompsoc
  \section*{Acknowledgments}
\else
  \section*{Acknowledgment}
\fi

The authors would like to thank...

\ifCLASSOPTIONcaptionsoff
  \newpage
\fi



%\bibliographystyle{IEEEtran}
%\begin{thebibliography}{1}
%\end{thebibliography}
%%-------------------------------------------------------------------------------

\bibliographystyle{IEEEtran}
\bibliography{IEEEexample}

%\begin{biography}[{\includegraphics[width=1in,height=1.25in,clip,keepaspectratio]{mshell}}]{Michael Shell}

\begin{IEEEbiography}{Rourab Paul}
Biography text here.
\end{IEEEbiography}

\begin{IEEEbiography}{Amlan Chakrabarti}
Biography text here.
\end{IEEEbiography}
\begin{IEEEbiography}{Ranjan Ghosh}
Biography text here.
\end{IEEEbiography}

%\newpage



%\vfill


\end{document}